\documentclass[]{jfm}
\pdfoutput=1
\usepackage{hyperref,lineno,amsmath,bm,amssymb,graphicx,multirow,algorithmic,color,nicefrac,microtype}
\hypersetup{colorlinks=true,citecolor=Violet,linkcolor=Red,urlcolor=Blue}

\definecolor{White}{rgb}{1.0,1.0,1.0}

\definecolor{Blue}{rgb}{0.12,0.05,0.8}

\renewcommand{\vec}[1]{\mathbf{#1}}

\newcommand{\vv}{\vec{v}}
\newcommand{\vvedge}{\vv_\text{e}}
\newcommand{\nor}{\vec{n}}
\newcommand{\vx}{\vec{x}}
\newcommand{\vX}{\vec{X}}

\newcommand{\vF}{\vec{F}}
\newcommand{\Ident}{\mathbf{1}}
\newcommand{\emach}{\epsilon_m}
\newcommand{\vxi}{\boldsymbol\xi}
\newcommand{\adv}{\text{adv}}
\newcommand{\istate}{\boldsymbol\zeta}
\newcommand{\vsig}{\boldsymbol\tau}
\newcommand{\vchi}{\boldsymbol\chi}
\newcommand{\gener}{a}
\newcommand{\alpad}{\alpha_\text{pad}}
\newcommand{\bepad}{\beta_\text{pad}}
\newcommand{\Trans}{\mathsf{T}}
\newcommand{\smov}[2]{Supplemental Movie~#1}
\renewcommand{\Re}{\textrm{Re}}

\usepackage{hyperref,color}
\definecolor{webgreen}{rgb}{0,.35,0}
\definecolor{webbrown}{rgb}{.6,0,0}
\definecolor{RoyalBlue}{rgb}{0,0,0.9}

\hypersetup{
   colorlinks=true, linktocpage=true, pdfstartpage=3, pdfstartview=FitV,
   breaklinks=true, pdfpagemode=UseNone, pageanchor=true, pdfpagemode=UseOutlines,
   plainpages=false, bookmarksnumbered, bookmarksopen=true, bookmarksopenlevel=1,
   hypertexnames=true, pdfhighlight=/O,
   urlcolor=webbrown, linkcolor=RoyalBlue, citecolor=webgreen,
   pdfauthor={Chris H. Rycroft},
   pdftitle={Reference Map Technique for Incompressible Fluid–Structure Interaction},
   pdfkeywords={},
   pdfcreator={pdfLaTeX},
   pdfproducer={LaTeX with hyperref}
}

\DeclareMathOperator{\sign}{sign}
\DeclareMathOperator{\Tr}{tr}

\shorttitle{Incompressible Reference Map Technique}
\shortauthor{C.~H.~Rycroft, C.-H. Wu, Y.~Yu, and K.~Kamrin}

\title{Reference Map Technique for Incompressible Fluid--Structure Interaction}
\author{Chris H.~Rycroft\aff{1,2}
\corresp{\email{chr@seas.harvard.edu}}, Chen-Hung Wu\aff{1,3}, Yue Yu\aff{4}, \and Ken Kamrin\aff{3}\corresp{\email{kkamrin@mit.edu}}}

\affiliation{
\aff{1}Paulson School of Engineering and Applied Sciences, Harvard University, Cambridge, MA~02139, USA
\aff{2}Mathematics Group, Lawrence Berkeley National Laboratory, 1 Cyclotron Road, Berkeley, CA~94720, USA
\aff{3}Department of Mechanical Engineering, Massachusetts Institute of Technology, Cambridge, MA~02139, USA
\aff{4}Department of Mathematics, Lehigh University, Bethlehem, PA~18015, USA
}

\begin{document}
\maketitle

\begin{abstract}
We present a general simulation approach for fluid--solid interactions based on
the fully-Eulerian Reference Map Technique (RMT). The approach permits the
modeling of one or more finitely-deformable continuum solid bodies interacting
with a fluid and with each other. A key advantage of this approach is its ease
of use, as the solid and fluid are discretized on the same fixed grid, which
greatly simplifies the coupling between the phases. We use the method to study
a number of illustrative examples involving an incompressible Navier--Stokes
fluid interacting with multiple neo-Hookean solids. Our method has several
useful features including the ability to model solids with sharp corners and
the ability to model actuated solids. The latter permits the simulation of
active media such as swimmers, which we demonstrate. The method is validated
favorably in the flag-flapping geometry, for which a number of experimental,
numerical, and analytical studies have been performed. We extend the flapping
analysis beyond the thin-flag limit, revealing an additional destabilization
mechanism to induce flapping.
\end{abstract}

\section{Introduction}
Fluid--structure interaction (FSI) problems highlight a natural dichotomy in
the simulation approaches for solids and fluids, where fluid problems tend to
be solved using Eulerian-frame methods
\citep{chorin67,hirt74,versteeg95,tannehill97} and solids with Lagrangian
approaches \citep{zienkiewicz67,sulsky94,hoover06,belytschko13}. An FSI
simulation method must therefore bridge the gap between these two perspectives.
For example, one set of FSI approaches treats both fluid and solid phases in a
Lagrangian frame, with a finite-element representation in the solid and an
adaptive Lagrangian mesh in the fluid
\citep{rugonyi01,bathe07,froehle15}, or with both phases treated with a
mesh-free approach \citep{rabczuk10}. An alternative methodology is to treat
the fluid on a fixed Eulerian mesh and the solid with Lagrangian points, such
as the family of immersed boundary methods
\citep{peskin02,griffith09,fai13}.

A fully Eulerian method whereby fluid and solid are both computed on a fixed
grid has its advantages. Computation time benefits arise from both phases being
treated on a single fixed background grid. The handling of multiple objects
interacting or of topological changes to objects can be done with level set
fields \citep{sethian99,osher88} rather than requiring complex on-the-fly
Lagrangian remeshing. In addition, certain common conditions such as
incompressibility are easier to implement in an Eulerian form. Lastly,
fixed-grid approaches are well-suited to numerical analysis, such as a von
Neumann stability analysis \citep{leveque_fd}.

The key challenge for a fully-Eulerian FSI method is to develop an Eulerian
description of the solid. In a small strain limit, this can be achieved by
writing the equations of linear elasticity in rate form, referred to as
hypoelasticity \citep{truesdell55}, which has formed the basis of several
numerical techniques \citep{udaykumar03,rycroft12,rycroft15}. However, here our
interest is in developing a large-deformation description of the solid, the
more general approach in solid mechanics \citep{gurtin10,belytschko13}. In
recent years, we have addressed the issue by developing an Eulerian-frame solid
simulation approach called the Reference Map Technique (RMT) \citep{kamrin09,
kamrinjmps, valkov15}, which is based on tracking the reference map
field---\textit{i.e.}~where material started from---on the Eulerian grid. The
reference map field allows the finite deformation of the solid to be computed,
from which the material stress is calculated according to a prescribed
nonlinear constitutive law. This approach has shown the ability to simulate
basic FSI and separately cover a span of desirable features. However, a single
implementation covering all needed features for robust physical
simulation---\textit{e.g.}~(i) numerical stability, (ii) second-order accuracy
in space and time, and (iii) desirable physical traits such as the ability to
model incompressible materials, objects with sharp corners, and activated
media---has been lacking and non-trivial to produce. In this paper we present
such a method and provide a variety of physical simulations using it, which
extend our understanding of certain FSI problems.

%To achieve these goals one must pose an Eulerian scheme capable of solving finite-deformation solid problems. The recently proposed Reference Map Technique (RMT) is such an Eulerian framework, based on tracking the \emph{reference map} field~\cite{kamrin09, kamrinjmps}. Other approaches for solid deformation on a fixed Eulerian grid include hypoelastic implementations~\cite{udaykumar03,rycroft12}, which may succeed for small elastic strains but lack a thermodynamically consistent form as needed for large deformations, and methods that directly evolve the deformation gradient tensor field as the primitive kinematic grid variable~\cite{plohr88,trangenstein91,liu01}. In a previous paper~\cite{kamrinjmps}, the RMT demonstrated the capability of accurately solving hyperelastic solid deformation problems on a fixed mesh---including shock propagation problems and problems with varied boundary and initial conditions---up to second-order accuracy in space and time. It also provided the first demonstrations of using the method to solve fully-coupled problems of FSI.

To represent incompressible solids and fluids we have reformulated the
numerical discretization using the projection method framework of
\citet{chorin67,chorin68}. In this method, to integrate the velocity field
forward by a time step, an intermediate velocity field is computed where the
incompressibility constraint is temporarily relaxed. After this, a Poisson
problem is solved for the pressure, which is used to project the velocity to be
divergence-free. The method has been extensively developed since Chorin's
original work \citep{brown01}. Here, we consider a modern second-order
implementation described by \citet{yu03,yu07} in the context of inkjet printer
nozzle simulation. This implementation incorporates a number of improvements,
including the treatment of advective terms by \citet{bell89}, and the
approximate projection approach of \citet{almgren96} based on a finite-element
discretization. We deliberately keep the fluid component of the simulation to
match this existing implementation, to emphasize that the reference map
technique does not require any special treatment of the fluid. However, we show
that the discretization techniques can be generalized to simulate solids via
the RMT, and we find that the advective discretization is especially
well-suited to simulating the reference map update equations in a fashion more
accurate than \citet{valkov15}.

The projection method removes the Courant--Friedrichs--Lewy (CFL) condition
\citep{courant67} associated with pressure waves. This makes it possible to
simulate a wide variety of problems in an intermediate Reynolds number regime
(and potentially for high Reynolds problems should an adaptive background grid
be used). As in \citet{valkov15}, the level set field representing interface(s)
is not explicitly updated, but is tied to where the boundary should be in the
reference map field. However, here we switch to a regression-based
extrapolation method, which is more stable, simpler, and allows shapes with
corners to be considered. Some accuracy tests are provided, demonstrating
second-order spatio-temporal accuracy. As a further test of this method for
physical simulation, we consider the flag flapping stability problem, which has
been studied extensively \citep{zhang00,watanabe02,zhu02,connell07}. We are
able to quantitatively reproduce the phase plot of stability for a thin flag
\citep{connell07} with very good accuracy for Reynolds numbers below 1000, and
reasonably good accuracy for Reynolds numbers above 1000. Our method also makes
it possible to simulate flags with substantial thickness, which show a
different instability mechanism due to vortex shedding from the tip. The
transition between the thin and thick flag behaviors is captured and studied
with our method. We also augment the approach to allow internal actuation of
the solid bodies. With this addition, the method is well-suited to
biolocomotion problems and we show an example of this tool by modeling a
jellyfish-like swimmer. Another advantage of the method is the ability to
perform many-body contact problems quickly but in a fashion that balances
momentum carefully. We demonstrate this approach with an example of many
objects of various sizes settling under gravity.

\section{Theory}
\label{sec:theory}
\subsection{Overview of the reference map technique}
We begin by considering the solid material, which we model using the
large-deformation hyperelastic framework~\citep{lubliner08,gurtin10}. As shown
in Fig.~\ref{fig:overview}(a), we introduce an undeformed reference
configuration for the solid at time $t=0$ with coordinate system $\vX$. We then
consider a time-dependent map $\vchi(\vX,t)$ from the undeformed
configuration to the deformed state in the physical frame at time $t$. The
deformation gradient tensor is defined as
\begin{equation}
  \label{eq:defrate}
  \vF = \frac{\p \vchi}{\p \vX}
\end{equation}
and represents how an infinitesimal element of the solid has been deformed and rotated.
From here, a consititutive law
\begin{equation}
  \label{eq:constit}
  \boldsymbol{\sigma}_s= \vec{f}(\vF,\istate)
\end{equation}
can be used to calculate the Cauchy stress $ \boldsymbol{\sigma}_s$ in the
physical frame, where $\istate$ represents any internal state variables such as
plastic deformation. The material velocity $\vv(\vx,t)$ then satisfies
\begin{equation}
  \label{eq:newton}
  \rho \left( \frac{\p \vv}{\p t} + (\vv \cdot \nabla) \vv \right) = \nabla \cdot  \boldsymbol{\sigma}
\end{equation}
where $ \boldsymbol{\sigma}= \boldsymbol{\sigma}_s$ in this case, $\rho =
\rho_s/(\det \vF)$, and $\rho_s$ is the solid density in the undeformed
configuration.

\setlength{\unitlength}{0.0025\textwidth}
\begin{figure}
  \begin{center}
    \scriptsize
    \begin{picture}(400,144)
      \put(0,0){\includegraphics[width=\textwidth]{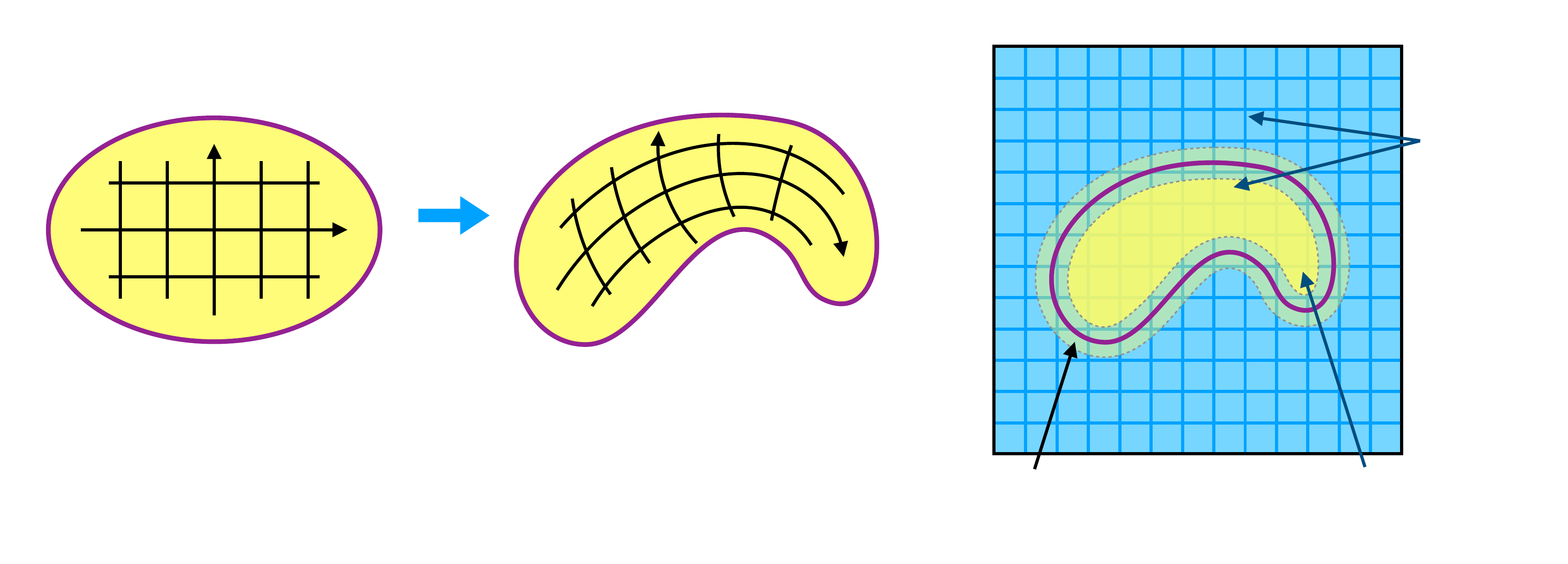}}
      \put(6,128){\footnotesize (a)}
      \put(236,128){\footnotesize (b)}
      \put(58,105){$\vX$}
      \put(78,106){\parbox{9em}{\centering Mapping \\ $\vchi(\vX,t)$}}
      \put(5,38){\parbox{12em}{\centering Initial undeformed \\ configuration}}
      \put(128,38){\parbox{12em}{\centering Deformed configuration \\ at time $t$}}
      \put(230,18){\textbf{Blur zone, $|\phi|<{w_T}$}}
      \put(315,14){\parbox{10em}{\centering \textbf{Solid reference} \\ \textbf{map, $\vxi(\vx,t)$}}}
      \put(336,107){\parbox{10em}{\centering \textbf{Global} \\ \textbf{velocity} \\ \textbf{field,} \\ $\vv(\vx,t)$}}
      \put(285,86){\textbf{Solid, $\phi<0$}}
      \put(265,117){\textbf{Fluid, $\phi>0$}}
    \end{picture}
  \end{center}
  \caption{(a) Overview of the hyperelastic framework, whereby an initially
  undeformed solid with reference coordinate system $\vX$ undergoes a
  time-dependent mapping $\vchi(\vX,t)$ to its current configuration at time
  $t$. (b) Overview of the reference map technique for simulating
  fluid--structure interaction on a fixed background grid. The sign of the
  level set function $\phi(\vx,t)$ demarcates the solid and fluid phases. The
  blur zone, used to transition from solid to fluid stress, is defined as the
  region where $|\phi|<w_T$.\label{fig:overview}}
\end{figure}
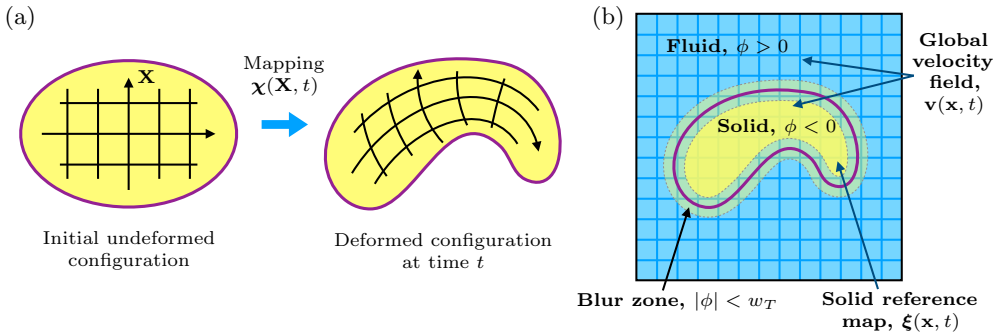

The most commonly used approach to simulate hyperelastic solids is to introduce
a deforming mesh on the solid, and then solve for the nodal displacements, from
which \eqref{eq:defrate} can be used to compute the
stress~\citep{belytschko13}. However, here we take an alternative approach of
introducing the \textit{reference map} field in the physical frame
$\vxi(\vx,t)$ that represents the inverse mapping of $\vchi$. The field is
initialized as $\vxi(\vx,0)=\vx$, and satisfies the advection equation
\begin{equation}
  \label{eq:rmap}
  \frac{\p \vxi}{\p t} + (\vv \cdot \nabla) \vxi = \vec{0}.
\end{equation}
The deformation gradient tensor is computed from the reference map field according to
\begin{equation}
  \label{eq:defrate2}
  \vF = \left( \frac{\p \vxi}{\p \vx} \right)^{-1},
\end{equation}
from which the Cauchy stress is evaluated. Equations \eqref{eq:constit},
\eqref{eq:newton}, \eqref{eq:rmap}, \& \eqref{eq:defrate2} then form a minimal
system of equations for finite-strain hyperelasticity in an Eulerian frame. The
reference map $\vxi(\vx,t)$ and velocity $\vv(\vx,t)$ can be represented on a
fixed grid. At each timestep equations \eqref{eq:defrate2} and
\eqref{eq:constit} can be used to evaluate the Cauchy stress, after which
equations \eqref{eq:newton} and \eqref{eq:rmap} can be integrated forward in
time. So far, this prescription is general, and could be solved using a variety
of discretization approaches such as a finite difference method, finite volume
method, or a discontinuous Galerkin method.

The reference map is a standard definition in solid mechanics \citep{gurtin10},
and it has been used in problems of inverse design
\citep{govindjee96,fachinotti08}, but it is not widely employed as a primary
simulation variable in the physical frame. Fixed-grid approaches by
\citet{plohr88,trangenstein91,liu01} have been developed that use the
deformation gradient tensor $\vF$ as a primary simulation variable.

\subsection{Incompressible fluid--structure interaction}
In this paper we employ the reference map technique to simulate incompressible
fluid--structure interactions. We shall use the terms $\vsig$, $\vsig_s$, and
$\vsig_f$ to refer only to the deviatoric part of the stress, as the pressure
field is now deformation independent and separately calculated. We make use of
a globally defined velocity field $\vv(\vx,t)$ that satisfies the
incompressibility constraint
\begin{equation}
  \label{eq:incomp}
  \nabla \cdot \vv = 0.
\end{equation}
We consider a solid immersed within the fluid, and introduce a level set
function $\phi(\vx,t)$ \citep{sethian,osher} that is the signed distance to the
solid--fluid interface with the convention that $\phi<0$ in the solid and
$\phi>0$ in the fluid. The reference map $\vxi(\vx,t)$ is defined within the
solid region only.

Let the fluid have density $\rho_f$ and dynamic viscosity $\mu$. The fluid stress
deviator at any gridpoint is given by
\begin{equation}
  \vsig_f = \mu ( \nabla \vv + (\nabla \vv)^\Trans).
\end{equation}
Kinematic viscosity is defined as $\nu=\mu/\rho_f$. The deviatoric stress is then
defined as a smooth transition between the fluid and solid stresses,
\begin{equation}
  \label{eq:stress_mix}
  \vsig = \vsig_s + H_\epsilon(\phi) (\vsig_f-\vsig_s),
\end{equation}
where
\begin{equation}
  \label{eq:sm_heavi}
  H_\epsilon(\phi) =
  \left\{
  \begin{array}{ll}
    0 & \qquad \text{if $\phi\le-\epsilon$,} \\
    \frac{1}{2} (1+\frac{\phi}{\epsilon}+\frac{1}{\pi} \sin \frac{\pi \phi}{\epsilon}) & \qquad \text{if $|\phi|<\epsilon$,} \\
    1 & \qquad \text{if $\phi\ge\epsilon$},
  \end{array}
  \right.
\end{equation}
is a smoothed Heaviside function with a transition region of width $2\epsilon$.
The detailed form of $H_\epsilon$ is not important, but the choice in
\eqref{eq:sm_heavi} has been used elsewhere \citep{yu03,yu07} and is twice
differentiable. In order to calculate $\vsig$ it is necessary to smoothly
extend $\vxi$ in the region $0<\phi<\epsilon$, which is done using
extrapolation methods that will be described in the following section. The
density is also defined as a smooth transition between the solid and fluid, as
\begin{equation}
  \label{eq:rho_mix}
  \rho = \rho_s + H_\epsilon(\phi) (\rho_s-\rho_f).
\end{equation}

\section{Numerical Method}
\label{sec:numerics}
The numerical procedure is based on the projection method of
\citet{chorin67,chorin68} for solving the incompressible Navier--Stokes
equations. Specifically, we consider a modern second-order method described by
\citet{yu03,yu07} that is especially effective at dealing with advection, and
incorporates a number of algorithmic advancements from Chorin's original
treatment.

\begin{figure}
  \begin{center}
    \footnotesize
    \begin{picture}(340,122)
      \put(0,0){\includegraphics[width=0.85\textwidth]{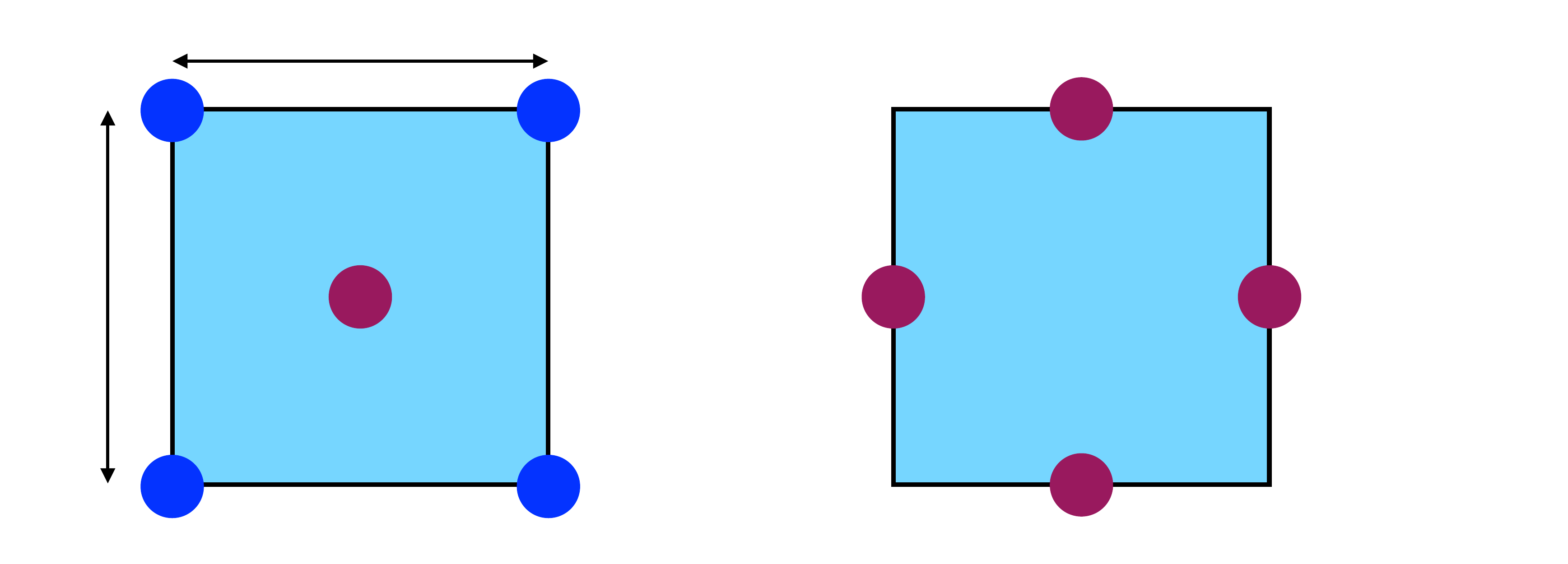}}
      \put(8,110){(a)}
      \put(167,110){(b)}
      \put(10,58){$h_y$}
      \put(74,113){$h_x$}
      \put(42,5){$p_{i,j}$}
      \put(42,87){$p_{i,j+1}$}
      \put(124,5){$p_{i+1,j}$}
      \put(124,87){$p_{i+1,j+1}$}
      \put(83,46){$\vv_{i,j},\vxi_{i,j}$}
      \put(83,33){$(\phi_{i,j})$}
      \put(198,39){\parbox{8em}{$\vv_{i-1/2,j},$ \\ $\vxi_{i-1/2,j}$}}
      \put(280,46){$\vv_{i+1/2,j},\vxi_{i+1/2,j}$}
      \put(239,5){$\vv_{i,j-1/2},\vxi_{i,j-1/2}$}
      \put(238,79){\parbox{8em}{$\vv_{i,j+1/2},$ \\ $\vxi_{i,j+1/2}$}}
    \end{picture}
  \end{center}
  \caption{(a) Arrangement of the fields within a simulation grid cell. The
  reference map $\vxi_{i,j}$, velocity $\vv_{i,j}$, and level set field
  $\phi_{i,j}$ are held at the cell center, while the pressure is held at the
  cell corners. The level set field is bracketed to emphasize that it is not
  time-evolved independently, but is instead derived from the reference map.
  (b) Arrangement of the edge velocities and reference maps that are computed
  at the half-timestep to evaluate the advective terms.\label{fig:grid}}
\end{figure}
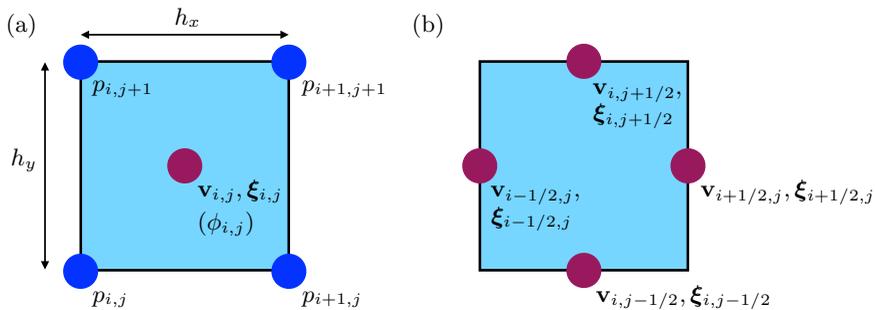

The simulation domain is a rectangle of size $W$ by $H$, and is divided into an
$M\times N$ grid of rectangular cells of size $h_x$ by $h_y$. The velocity, the
reference map, and the level set, are held at cell centers and are indexed as
$\vv_{i,j}$, $\vxi_{i,j}$, and $\phi_{i,j}$, respectively, for $i=0,\ldots,
M-1$ and $j=0,\ldots, N-1$ (Fig.~\ref{fig:grid}(a)). The components of the
velocity field are written as $\vv_{i,j}=(u_{i,j},v_{i,j})$. Pressures are held
at cell corners and are indexed as $p_{i,j}$ for $i=0,\ldots, M$ and
$j=0,\ldots, N$. In addition, the grid is padded by two layers of cells in each
direction whose values are populated to enforce different boundary conditions.

Superscripts are used to denote timesteps. To advance the simulation forward
from timestep $n$ to $n+1$ with interval $\Delta t$, the following procedure
is used. The reference map field is updated using
\begin{equation}
  \frac{\vxi^{n+1}-\vxi^n}{\Delta t} = -\left[ (\vv \cdot \nabla) \vxi\right]^{n+1/2}
  \label{eq:main_rmap}
\end{equation}
and an intermediate velocity $\vv^*$ is computed using
\begin{equation}
  \label{eq:main_inter}
  \frac{\vv^{*}-\vv^n}{\Delta t} = -\left[(\vv\cdot \nabla)\vv\right]^{n+1/2}
  + \frac{\mu}{\rho(\phi^{n+1/2})} \nabla \cdot \left[\vsig(\vxi^{n+1/2},\vv^n)\right].
\end{equation}
Here, the advective derivatives $[(\vv\cdot \nabla) \vxi]^{n+1/2}$ and $[(\vv
\cdot \nabla) \vv]^{n+1/2}$ are evaluated at the middle of the timestep using a
second-order explicit Godunov scheme, described in Subsec.~\ref{sub:advection}.
Once the advective derivatives are evaluated, Eq.~\eqref{eq:main_rmap} allows
$\vxi^{n+1}$ to be computed. This allows the time-centered reference map to be
defined as $\vxi^{n+1/2} = (\vxi^n+\vxi^{n+1})/2$ after which $\vv^{*}$ is
computed using Eq.~\eqref{eq:main_inter}. From here, the Poisson problem for
pressure is evaluated using
\begin{equation}
  \label{eq:main_poiss}
  \nabla \cdot \vv^* = \nabla\cdot\left( \frac{\Delta t}{\rho(\phi^{n+1/2})} \nabla p^{n+1}\right).
\end{equation}
Following \citet{almgren96,puckett97}, Eq.~\eqref{eq:main_poiss} is solved using
a finite-element formulation, described in Subsec.~\ref{sub:finite_element}.
After this, the velocity is projected to be divergence-free using
\begin{equation}
  \label{eq:main_proj}
  \vv^{n+1} = \vv^* - \frac{\Delta t}{\rho(\phi^{n+1/2})} \nabla p^{n+1}
\end{equation}
where the gradient of $p^{n+1}$ is evaluated using a second-order centered
difference formula.

\subsection{Advective terms}
\label{sub:advection}
To evaluate the advective terms appearing in Eqs.~\eqref{eq:main_rmap} and
\eqref{eq:main_inter}, a second-order explicit Godunov scheme is used. The same
scheme is applied to both the velocity $\vv$ and reference map $\vxi$.
Throughout this section, we denote $\gener$ to be a generic scalar component of
either of these two fields. We also refer the reader to recent work by
\cite{jain17}, which introduces an alternative numerical treatment for
reference map advection.

\subsubsection{Godunov upwinding}
\label{sss:godunov}
To begin, the gradients of the reference map and velocity at each cell center
are evaluated using the fourth-order monotonicity-limited scheme of
\citet{colella85} described in Appendix \ref{app:mono_deriv}. Once the
gradients are calculated at the center of each cell, edge-centered velocities
and reference maps are created at $t+\Delta t/2$ using Taylor expansions to
each of the four edges, which are indexed using half-integers as shown in
Fig.~\ref{fig:grid}(b). As an example, an extrapolation of the reference map to
the right edge of the cell (with superscript $R$) is given by
\begin{align}
  \label{eq:godunov1}
  \vxi^{R,n+1/2}_{i+1/2,j} &= \vxi_{i,j}^n + \frac{\Delta t}{2} (\p_t \vxi)^n_{i,j} + \frac{h_x}{2} (\p_x \vxi)^n_{i,j} \nonumber \\
  &= \vxi_{i,j} + \frac{1}{2} \left( h_x - u^n_{i,j} \Delta t\right) \vxi^n_{x,i,j} - \frac{\Delta t}{2} (\widetilde{v \vxi_y})^n_{i,j},
\end{align}
where Eq.~\eqref{eq:rmap} has been substituted for the $\vxi_t$ derivative. The
extrapolation of the velocity to the right edge is
\begin{align}
  \label{eq:godunov2}
  \vv^{R,n+1/2}_{i+1/2,j} &= \vv^n_{i,j} + \frac{\Delta t}{2} \vv^n_{t,i,j} + \frac{h_x}{2} \vv^n_{x,i,j} \nonumber \\
  &= \vv^n_{i,j} + \frac{1}{2}\left(h_x - u^n_{i,j} \Delta t\right) \vv^n_{x,i,j} - \frac{\Delta t}{2} (\widetilde{v \vv_y})^n_{i,j} - \frac{\Delta t}{2} \vec{a}^n_{i,j},
\end{align}
where
\begin{equation}
  \label{eq:godunov_force}
  \vec{a}^n_{i,j} = \left[ -\frac{1}{\rho} \nabla p + \frac{1}{\rho(\phi)} \nabla \cdot \vsig \right]_{i,j}.
\end{equation}
Equivalent procedures are used to compute extrapolations left, down, and up
with superscripts $L$, $D$, and $U$, respectively. To ensure tangential
stability the terms with tildes in Eqs.~\eqref{eq:godunov1} \&
\eqref{eq:godunov2} are computed differently using the procedure in Appendix
\ref{app:tang}. After this procedure, each edge has velocities and reference
maps from the two cells that adjoin it, and a Godunov upwinding procedure is
used to select which values to use. On the vertical edge at
$(i+\nicefrac{1}{2},j)$,
\begin{equation}
  \label{eq:gselect}
  \gener^{n+1/2}_{i+1/2,j}=
  \begin{cases}
    \gener^{L,n+1/2}_{i+1/2,j} & \qquad \text{if $u^{L,n+1/2}_{i+1/2,j}>0$ and $u^{L,n+1/2}_{i+1/2,j}+u^{R,n+1/2}_{i+1/2,j}>0$,} \\
    \gener^{R,n+1/2}_{i+1/2,j} & \qquad \text{if $u^{R,n+1/2}_{i+1/2,j}<0$ and $u^{L,n+1/2}_{i+1/2,j}+u^{R,n+1/2}_{i+1/2,j}<0$,} \\
    \mathcal{F}(\gener^{L,n+1/2}_{i+1/2,j},\gener^{R,n+1/2}_{i+1/2,j}) & \qquad \text{otherwise.}
  \end{cases}
\end{equation}
where $\gener$ is a generic component. Thus if the velocity field points
rightward then the components are taken from the left cell, and if the velocity
field points leftward then the components are taken from the right cell. The
function $\mathcal{F}$ is used when the two velocities are ambiguous. For the
horizontal velocity $\mathcal{F}(\beta,\gamma)=0$ (Case A), and for all other
components $\mathcal{F}(\beta,\gamma)=(\beta+\gamma)/2$ (Case B). On an edge
where a velocity boundary condition is applied (\textit{e.g.}\ a no-slip
condition) the corresponding edge velocity is set to exactly match the
condition. In this paper we restrict to cases of localized solid objects that
do not extend to the boundary and thus we do not apply special boundary
condition treatment for edge reference map fields.

\subsubsection{Marker-and-cell (MAC) projection}
\label{sss:mac_proj}
The edge velocities calculated in Subsubsec.~\ref{sss:godunov} may not be
precisely divergence free. We therefore apply an intermediate MAC projection
step to ensure that the discrete flux entering any grid cell is exactly zero.
Let $\vvedge$ be the edge velocities, and let $q$ be a cell-centered scalar
field. We aim to make
\begin{equation}
  \label{eq:mac1}
  \vvedge - \frac{1}{\rho} \nabla q
\end{equation}
divergence free. Taking the divergence of Eq.~\eqref{eq:mac1} yields
\begin{equation}
  \label{eq:mac2}
  \nabla\cdot \left(\frac{1}{\rho} \nabla q\right) = \nabla \cdot \vvedge,
\end{equation}
which is discretized as
\begin{multline}
  \label{eq:mac3}
  \frac{1}{h_x^2} \left( \frac{q_{i+1,j} - q_{i,j}}{\rho_{i+1/2,j}}
  + \frac{q_{i,j}-q_{i-1,j}}{\rho_{i-1/2,j}} \right)
  + \frac{1}{h_y^2} \left( \frac{q_{i,j+1}-q_{i,j}}{\rho_{i,j+1/2}}
  + \frac{q_{i,j}-q_{i,j-1}}{\rho_{i,j-1/2}} \right) \\
  = \frac{u_{i+1/2,j} - u_{i-1/2,j}}{h_x} + \frac{v_{i,j+1/2}-v_{i,j-1/2}}{h_y}.
\end{multline}
Edge-based densities appearing in this equation are computed via linear
interpolation from the two adjacent grid cells. At boundaries where a velocity
boundary condition is applied, any derivative on the left hand side of
Eq.~\eqref{eq:mac3} is omitted if it contains $q$ values that are out of range.
If a pressure condition is applied, then a Dirichlet condition of $q=\Delta
t\,p/2$ is applied, where the factor of two arises because the edge velocities
are time-centered.

Equation \eqref{eq:mac3} results in a large linear system $Aq=b$ where $A$ is a
sparse matrix, $b$ is the source term, and $q$ is a vector of the components
$q_{i,j}$. This is solved with a custom multigrid C++ library that employs
multithreading using OpenMP. Since the $q$ field typically varies smoothly in
time, the initial guess for the multigrid algorithm is computed as a linear
interpolation from the previous two timesteps. Multigrid V-cycles are performed
until the mean squared element in the residual vector $r=Aq-b$ reaches a
required tolerance $T_\text{MAC}$. We assume that velocities and densities are
within several orders of magnitude of unity. Then an appropriate scale for an
element of the residual is $r_\text{s}= 4(h_x^{-2}+h_y^{-2})\Delta t$, and a
tolerance of $T_\text{MAC}=10^{4} r_s \emach$ is used, where $\emach$ is the
machine epsilon for double precision floating point arithmetic. Once the
tolerance level is reached, one further V-cycle is performed to further improve
accuracy. Typically 5--15 V-cycles are required.

\subsubsection{Evaluation of the derivative}
\label{sss:deriv_eval}
Once the MAC projection has been performed the time-centered advective term for
the velocity and reference maps are evaluated as
\begin{align}
  \left[ (\vv \cdot \nabla) \gener\right]^{n+1/2}_{i,j} &=
  \frac{u^{n+1/2}_{i+1/2,j} + u^{n+1/2}_{i-1/2,j}}{2} \frac{
  \gener^{n+1/2}_{i+1/2,j} - \gener^{n+1/2}_{i-1/2,j}}{h_x} \nonumber \\
  &\phantom{=} +
  \frac{v^{n+1/2}_{i,j+1/2} + v^{n+1/2}_{i,j-1/2}}{2} \frac{
  \gener^{n+1/2}_{i,j+1/2} - \gener^{n+1/2}_{i,j-1/2}}{h_y}
\end{align}
where $\gener$ is a generic field component.

\subsection{Level set update and reference map extrapolation}
The simulation makes use of a cell-centered level set function $\phi_{i,j}$ for
tracking the fluid--solid boundary. The level set routine is stored in a narrow
band of points of width $2\phi_W$ surrounding the interface
\citep{sethian,rycroft12} that is chosen to be large enough to contain the
entire blur zone and perform finite difference calculations at all points in
this region. The level set is used to extrapolate the reference map fields in
the narrow band, and to calculate the mixing of stress and density according to
Eqs.~\eqref{eq:stress_mix} \& \eqref{eq:rho_mix}, respectively. Unlike typical
applications of the level set method, the $\phi$ field is not explicitly
updated, but is instead continually given by the reference map field using the
procedure first described in \citet{valkov15}.

\subsubsection{Level set construction}
For a given shape, define a continuous function of the reference map
$\phi_0(\vxi)$ such that $\phi_0<0$ for reference map values in the solid,
$\phi_0>0$ for reference map values outside the solid, and $\phi_0=0$ on the
interface. During the timestep, the reference map field $\vxi^{n+1}$ is
computed inside the solid using Eq.~\eqref{eq:main_rmap}, from which the
half-timestep reference map is defined as $\vxi^{n+1/2}=(\vxi^n+\vxi^{n+1})/2$.
Both fields are extended into the narrow band fluid region using the
extrapolation methods described in Subsubsec.~\ref{sss:extrap}.

To construct the new level set function $\phi^{n+1/2}$, an auxiliary function
$\psi^{n+1/2}$ is first computed in the narrow band such that
$\psi^{n+1/2}=\phi_0(\vxi^{n+1/2})$. The zero contour of $\psi^{n+1/2}$ will
lie at the fluid--solid interface, but this function itself may not satisfy the
signed-distance property. To recover the signed-distance property, the level
set $\phi^{n+1/2}$ is constructed from $\psi^{n+1/2}$ using the
reinitialization procedure described in \citet{rycroft12}. This procedure first
considers points $(i,j)$ that straddle the interface, so that one of their
orthogonal neighbors has a \smash{$\psi^{n+1/2}$} value of an opposite sign.
Each straddling point is considered. The bicubic interpolant
\smash{$\psi^{n+1/2}_\text{c}$} is computed, and the modified Newton--Raphson
approach of \citet{chopp01,chopp09} is used to find the shortest distance
vector $\boldsymbol\Delta \vx$ from each straddling point to the interface
\smash{$\psi^{n+1/2}_\text{c}(\vx)=0$}, after which the level set function is
initialized to $\pm |\boldsymbol\Delta \vx|$. In extremely rare cases the
root-finding method can fail, in which case the routine falls back on the
first-order method described by \citet{sethian}. For further details, see
\citet{rycroft12}.

With the straddling points of $\phi^{n+1/2}$ now initialized, the remaining
points are filled in using the second-order fast marching method of
\citet{sethian}. In the fluid, the positive $\phi^{n+1/2}$ values are computed
in order of increasing value, until reaching a cutoff $\phi_W$ that defines the
width of the narrow band. The same procedure is used to fill in the negative
$\phi^{n+1/2}$ values in the solid, until reaching a cutoff $-\phi_W$. After
this procedure, the level set function is now a signed-distance function inside
the narrow band. Note that these routines work reliably even if the function
$\psi_{n+1/2}$ has a loss of regularity as some points: since the entire
$\phi^{n+1/2}$ field is directly constructed, there is no possibility for
instabilities to grow over time, as can happen in PDE-based update procedures.
Identical methods are used to construct $\phi^{n+1}$ from $\vxi^{n+1}$.

\subsubsection{Extrapolation}
\label{sss:extrap}
During the construction of the level set function, a list of non-straddling
fluid points sorted in order of increasing value, $0<\phi_1<\phi_2<\ldots$ is
constructed, which is used for extrapolating the reference map $\vxi$ from the
solid into the fluid narrow band. Previous approaches to do this have employed
PDE-based methods by defining a normal vector $\nor=\nabla \phi$ and
extrapolating outwards from the object in the direction of $\nor$
\citep{aslam04,rycroft12}. While these methods are well-suited to mathematical
analysis, they require considerable bookkeeping for performing the finite
difference calculations of $\phi$ and $\vxi$ due to the fields only existing at
certain grid locations. In previous work we have found this to be a source of
difficulty \citep{valkov15}.

In the current work, we make use of the following alternative extrapolation
procedure. Consider the points in increasing order of $\phi$ value. For a
particular point $(i,j)$ at physical location $\vx_{i,j}$:
\begin{enumerate}
  \item Set $r=3$.
  \item Use least-squares regression to fit a linear map
    $\vxi_\text{lm}(\vx)=\vec{A}x+\vec{B}y+\vec{C}$ using all available
    reference map values at $(i',j')$ such that $|i-i'|\le r, |j-j'|\le r$.
    Weight each value in the regression according to $\phi_{i,j}-\phi_{i',j'}$.
  \item If the linear map is degenerate then increment $r$ and return to Step
    2. Otherwise, continue.
  \item Set $\vxi_{i,j}=\vxi_\text{lm}(\vx_{i,j})$.
\end{enumerate}
This procedure is simpler than the PDE-based methods since it does not require
extensive bookkeeping. Since the method uses all available values in a
neighborhood, this repeated averaging results in substantial blurring if the
extrapolation is continued far away from the interface. However, here, only
values near the interface are required, and the averaging is beneficial,
serving to damp out high-frequency modes that could be the source of
instability. In Step 3, degeneracies occur only when the available points are
colinear, in which case there is insufficient information to determine the
linear map. In this case, Step 4 causes the algorithm to retry using more
neighboring points.

\subsection{Computation of stress}
\label{sub:stress_comp}
In order to evaluate the stress divergence terms that appear in
Eq.~\eqref{eq:main_inter} \& \eqref{eq:godunov_force}, the stresses are first
computed on the edges of each grid cell. The stress term in
Eq.~\eqref{eq:godunov_force} is computed as
\begin{equation}
  \nabla \cdot \left[\vsig(\vxi^n)\right] = \frac{[\vsig_x]^n_{i+1/2,j}+[\vsig_x]^n_{i-1/2,j}}{h_x}
  + \frac{[\vsig_y]^n_{i,j+1/2}-[\vsig_y]^n_{i,j-1/2}}{h_y}
\end{equation}
where $\vsig_x=(\tau_{xx},\tau_{xy})$ and
$\vsig_y=(\tau_{xy},\tau_{yy})$ are the components acting on the vertical
and horizontal edges, respectively.

\subsubsection{Solid stress}
To begin, the components of the Jacobian are computed using the second-order
finite difference formulae
\begin{equation}
  \left(\frac{\p \vxi}{\p x}\right)_{i-1/2,j} = \frac{\vxi_{i,j}-\vxi_{i-1,j}}{h_x},
  \qquad \left(\frac{\p \vxi}{\p y}\right)_{i-1/2,j} = \frac{\vxi_{i,j+1}+\vxi_{i-1,j+1}
  -\vxi_{i,j-1}-\vxi_{i-1,j-1}}{4h_y}
\end{equation}
after which the deformation gradient is evaluated as
\begin{equation}
  \vF_{i-1/2,j} = \left( \left(\frac{\p \vxi}{\p \vx}\right)_{i-1/2,j} \right)^{-1}.
\end{equation}
From here, any constitutive law $\vsig_s=\vec{f}(\vF)$ could be used to
evaluate the deviatoric stress, $\vsig_s$. In the current work, we employ the
plane-strain incompressible neo-Hookean law,
\begin{equation}
  \label{eq:nsstress}
  \vsig_s = \vec{f}(\vF) = G \left( \vF \vF^\Trans - \tfrac{1}{3} \Ident (\Tr \vF \vF^\Trans +1)\right),
\end{equation}
where $G$ is the small-strain shear modulus.

\subsubsection{Fluid stress}
To evaluate the fluid stress, the gradients of the velocity on vertical edges
are computed as
\begin{align}
  \left(\frac{\p \vv}{\p x}\right)_{i-1/2,j} &= \frac{\vv_{i,j}-\vv_{i-1,j}}{h_x}, \label{eq:vgrad1} \\
  \left(\frac{\p \vv}{\p y}\right)_{i-1/2,j} &= \frac{\vv_{i,j+1}+\vv_{i-1,j+1}
  -\vv_{i,j-1}-\vv_{i-1,j-1}}{4h_y}. \label{eq:vgrad2}
\end{align}
Equivalent stencils are used to compute velocity gradients on horizontal edges,
after which the fluid stress is given by
\begin{equation}
  \label{eq:nfstress}
  \vsig_f= \mu (\nabla \vv + (\nabla \vv)^\Trans)
\end{equation}
where $\mu$ is the viscosity. Equation~\eqref{eq:nfstress} is our standard
approach for computing the fluid stress. However, we have also investigated
a simplified calculation. Since $\nabla \cdot \vv = 0$, it follows that in the
bulk of the fluid, the second term in Eq.~\eqref{eq:nfstress} has zero
contribution to $\nabla\cdot \vsig_f$. Hence an alternative formula is
\begin{equation}
  \label{eq:nfstress_alt}
  \vsig_f^\text{(simp)} = \mu \nabla \vv.
\end{equation}
This formula only requires evaluating the simpler stencil in
Eq.~\eqref{eq:vgrad1}. However, Eq.~\eqref{eq:nfstress_alt} is not strictly
valid in the blur zone since taking the divergence Eq.~\eqref{eq:stress_mix}
results in a non-zero contribution from the second term of
Eq.~\eqref{eq:nfstress_alt}.

Once all edge stresses are computed, the divergence is computed using
\begin{equation}
  \left[\nabla \cdot \vsig \right]_{i,j} = \frac{\vsig_{i+1/2,j}-\vsig_{i-1/2,j}}{h_x}
  + \frac{\vsig_{i,j+1/2}-\vsig_{i,j-1/2}}{h_y}.
\end{equation}

\subsection{Finite-element projection}
\label{sub:finite_element}
To solve the Poisson problem in Eq.~\eqref{eq:main_poiss}, we make use of a
finite-element formulation. The pressure is comprised of piecewise bilinear
elements, and the velocity and density are piecewise constant on the grid
cells. For a given pressure element $\psi$ the weak formulation of
Eq.~\eqref{eq:main_poiss} is
\begin{equation}
  \label{eq:fem}
  \int_\Omega \vv^* \cdot \nabla \psi \,dx\,dy - \int_\Omega \frac{\Delta t}{\rho(\phi^{n+1/2})} \nabla p^{n+1} \cdot \nabla \psi \,dx \,dy
  = \int_{\Gamma_1} \psi \vv^\text{BC} \cdot \vec{n}\, dS
\end{equation}
where $\Gamma_1$ is the section of the boundary where inflow and outflow
conditions are prescribed. Consider a particular bilinear element function
$\psi$ located at a pressure point $p_{i,j}$ in the bulk of the domain. The
first term of Eq.~\eqref{eq:fem} is
\begin{equation}
  h_x (u^*_{i+1,j+1} + u^*_{i+1,j} - u^*_{i,j+1} - u^*_{i,j})
  +h_y (v^*_{i+1,j+1} - v^*_{i+1,j} + v^*_{i,j+1} - v^*_{i,j})
\end{equation}
and the second term is
\begin{equation}
  \lambda_a p_{i,j} + \lambda_b(p_{i-1,j}+p_{i+1,j}) + \lambda_c (p_{i,j-1}+p_{i,j+1})
  +\lambda_d \sum_{\substack{k=\pm1 \\ l=\pm1}} p_{i+k,j+l}.
\end{equation}
where
\begin{equation}
  \lambda_a = \frac{4(h_x^2+h_y^2)}{3h_xh_y},\,\,\,
  \lambda_b = \frac{-2h_y^2+h_x^2}{3h_xh_y},\,\,\,
  \lambda_c = \frac{-2h_x^2+h_y^2}{3h_xh_y},\,\,\,
  \lambda_d = \frac{-h_x^2-h_y^2}{6h_xh_y}.\,\,\,
\end{equation}
This linear system is solved using the multithreaded custom C++ geometric
multigrid library mentioned in Subsubsec.~\ref{sss:mac_proj} using an error
tolerance of $T_\text{FEM}=\lambda_a 10^4 \emach$.

\subsection{Parameter choices and stability}
\label{sub:param_stab}
Our test cases involve four physical parameters: solid shear modulus $G$, solid
density $\rho_s$, fluid viscosity $\mu$, and fluid density $\rho_f$. In the
solid, the shear wave speed is \smash{$c_s=\sqrt{G/\rho_s}$}. The CFL condition
requires that the simulation timestep be less than or equal to
\begin{equation}
  \Delta t_\text{I} = c_s \min\{h_x,h_y\} = \sqrt{\frac{G}{\rho_s}} \min\{h_x,h_y\}.
  \label{eq:trestrict1}
\end{equation}
In addition, performing a von Neumann stability analysis shows that the
timestep must be less than or equal to
\begin{equation}
  \Delta t_\text{II} = \frac{\rho_f }{2\mu(h_x^{-2} + h_y^{-2})}
  \label{eq:trestrict2}
\end{equation}
in order to resolve the viscous fluid stress. Inside the solid, we find that
simulating stress using only Eq.~\eqref{eq:nsstress} results in an
instability---this should be expected since we are effectively solving a
hyperbolic system using centered finite differences. To rectify this, we
incorporate an extra small artificial viscous stress inside the solid. Based on
dimensional considerations, the artificial viscosity should satisfy
\begin{equation}
  \label{eq:evisc_scale}
  \mu_e = \kappa_e \rho_s c_s \max\{ h_x, h_y\}
\end{equation}
where $\kappa_e$ is a dimensionless constant. In addition, we also find that
artificial viscosity is useful in the fluid--solid transition region. We
therefore define the extra viscous stress as
\begin{equation}
  \vsig_e(\vx) = \mu_e \left[ (1-H(\phi(\vx))) + q (1-w_T H'(\phi(\vx)))\right] \nabla \vv.
\end{equation}
where $q$ is a dimensionless constant. Based on a variety of tests, we use
$q=1$ and $\kappa_e=0.4$ throughout the paper. Since the purpose of this extra
stress is to stabilize the numerical system, we employ the simpler form of
fluid stress given in Eq.~\ref{eq:nfstress_alt}. Since $\mu_e$ scales linearly
with grid spacing, and the simpler fluid stress only introduces a discrepancy
in the blur zone, any errors that are introduced will reduce to zero as the
grid is refined. The corresponding timestep restriction is
\begin{equation}
  \Delta t_\text{III} = \frac{\rho_s }{2 \mu_e (1+q) (h_x^{-2} + h_y^{-2})}.
  \label{eq:trestrict3}
\end{equation}
With these definitions in place, the simulation timestep $\Delta t$ is chosen
to be smaller than the minimum of the three conditions in
Eqs.~\eqref{eq:trestrict1}, \eqref{eq:trestrict2}, \& \eqref{eq:trestrict3}, so
that
\begin{equation}
  \Delta t = \min\{ \alpad \Delta t_\text{I},
  \alpad \Delta t_\text{II}, \bepad \Delta t_\text{III}\}.
\end{equation}
Here, $\alpad$ and $\bepad$ are padding factors that are smaller than one. For
this paper we use $\alpad=0.4$ and $\bepad=0.8$, so that the restrictions
arising from the two physical stresses (I \& II) are applied more stringently
than the one for the artificial stress (III). Note that in the limit as
$h_x,h_y\to 0$, the artificial viscosity vanishes.

\section{Results}
\label{sec:results}
Since our purpose is to demonstrate the numerical method as opposed to apply it
to a specific problem, we make use of non-dimensionalized quantities for all of
the results that we present. To connect the results to reality, the simulation
parameters and output can be multiplied by appropriate length, time, and mass
scales. Our results also focus on the case of equal grid spacing, $h_x=h_y=h$.

\subsection{A spinning flexible rotor}
\label{spin_rotor}
The first example demonstrates our method's ability to handle sharp solid
corners within a non-trivial FSI setting. It consists of a spinning flexible
regular seven-pointed star that is centered on the origin and has a vertex at
\smash{$(0.62\cos \tfrac{2\pi k}{7},\sin\tfrac{2\pi k}{7})$} for $k\in
\mathbb{Z}$, with density $\rho_s=3$. The resolution is $800\times 800$, the
simulation domain is $[-1,1)^2$ and periodic boundary conditions are used. The
fluid and rotor are initially stationary. The region $r=|\vx|<0.16$ is used as
a pivot. To excite the fluid, the pivot is rotated with an oscillatory motion
with angle $\theta(t)=\pi (1-\cos t)$. This is done by applying an external
tether force to the pivot region of
\begin{equation}
  \vec{f}_\text{teth}(\vx) = K_\text{teth} H_\epsilon(r-r_\text{teth}) (\vec{x}
  - R_{\theta(t)} \vxi(\vx))
  \label{eq:tether}
\end{equation}
where $r_\text{teth}=0.16$ and $R_{\theta(t)}$ is a rotation matrix with angle
$\theta(t)$. The spring constant is set to $K_\text{teth}=10^{-2} \rho_s \Delta
t^2$, which ensures that the natural frequency of the tether satisfies the
timestep restriction imposed by the method.

The simulation was run from $t=0$ to $t=4\pi$ using sixteen threads on a Linux
server with two Intel Xeon E5-2683 2.10~GHz processors. For the given
parameters, the timestep of $\Delta t=1.105\times 10^{-4}$ was determined by
the extra viscous stress in the solid. Simulation output was saved at regular
intervals of $\pi/150$. The total wall clock time for the simulation was
10.48~h. A total of 114,000 timesteps were performed, with each taking 0.33~s
to compute. A substantial fraction of the computation time is spent performing
the two linear solves. The MAC projections take on average 14.78 V-cycles and
require 0.08~s per timestep. The finite element projections take on average
13.48 V-cycles and require 0.11~s per timestep.

Snapshots of the simulation are shown in Fig.~\ref{fig:seven-star}. As the star
begins to rotate, each point deforms clockwise, and vortices are shed from the
points, which are visible at $t=4\pi/15$. By $t=\pi$, the rotor is stationary,
and the points are now deformed anti-clockwise due to the angular deceleration.
As time progresses, the disturbance to the fluid becomes larger. By $t=2\pi$,
the rotational symmetry of the fluid flow is lost, due to interactions across
the periodic boundaries, which break the seven-fold symmetry. By $t=4\pi$,
after two complete cycles of the oscillatory motion, there are vortices present
throughout the fluid. \smov{1}{seven-star} shows the complete simulation. To
visualize the fluid motion, the movie also shows a number of tracers with
trajectories $\vx(t)$. The tracers are initialized at random positions in the
fluid and are updated using the ordinary differential equation
\begin{equation}
  \frac{d\vx}{dt}=\vv_\text{bic}(\vx(t)),
  \label{eq:tracer}
\end{equation}
where $\vv_\text{bic}$ is the bicubic interpolation of the velocity field, and
the time integration is performed using the second-order improved Euler method
\citep{suli}.

\setlength{\unitlength}{0.0025\textwidth}
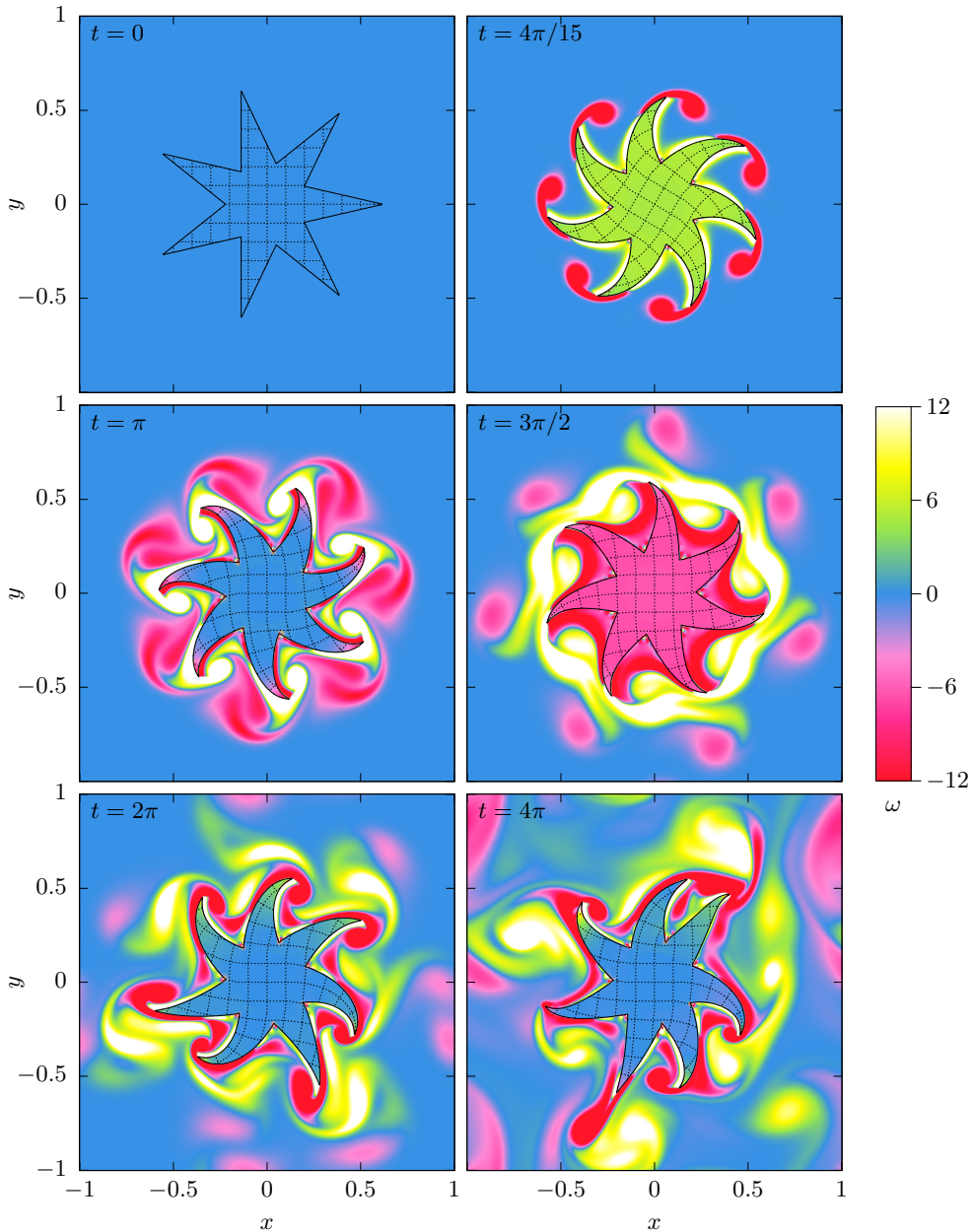
\begin{figure}
  \begin{center}
    \begin{picture}(400,510)
      \put(8,250){\parbox{\textwidth}{\include{seven-star_fig}}}
      \put(354,190){\includegraphics[width=14\unitlength,height=148\unitlength]{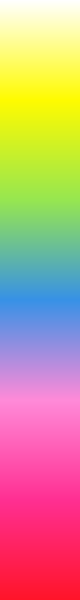}}
      \put(354,190){\line(0,1){148}}
      \put(368,190){\line(0,1){148}}
      \put(354,190){\line(1,0){18}}
      \put(354,338){\line(1,0){18}}
      \put(368,227){\line(1,0){4}}
      \put(368,264){\line(1,0){4}}
      \put(368,301){\line(1,0){4}}
	  \put(374,190){\makebox(0,0)[l]{$-12$}}
	  \put(374,227){\makebox(0,0)[l]{$-6$}}
	  \put(374,264){\makebox(0,0)[l]{$0$}}
	  \put(374,301){\makebox(0,0)[l]{$6$}}
	  \put(374,338){\makebox(0,0)[l]{$12$}}
      \put(361,180){\makebox(0,0)[c]{$\omega$}}
    \end{picture}
  \end{center}
  \vspace{-1em}
  \caption{Snapshots of vorticity $\omega$ in a simulation of flexible
  seven-pointed rotor being spun with an oscillatory motion a fluid. The thick
  black line marks the fluid--structure interface. The thin dashed lines are
  contours of the components of the reference map and indicate how the rotor
  has deformed.\label{fig:seven-star}}
\end{figure}

\subsection{Tests of convergence and accuracy}
To study the accuracy of the numerical method, we performed a convergence test
in the periodic domain $[-1,1)^2$ using an initial incompressible velocity
field of
\begin{equation}
  \vv(\vx,0) = \sum_{k=0}^5 (-1)^k \vv_\text{vor}\left(x-\frac{-5+2k}{6},y-\frac{-5+2k}{6},2(k+1)\right)
\end{equation}
where
\begin{equation}
  \vv_\text{vor}(\vx,\lambda) = ( -\sin \pi y, \cos \pi x) \times e^{-\lambda(2-\cos \pi x -\cos \pi y)}.
\end{equation}
This velocity field is designed to have features with a variety of length
scales. We simulated up to $t=0.5$, used a shear modulus of $G=1$, a fluid
density of $\rho_f=1$, and employed the standard choices for extra viscosity
and timestep selection. Using the same initial velocity field, we ran tests
using (i) fluid only, (ii) solid only, (iii) a circle of radius 0.6 centered on
$(-0.1,0)$, and (iv) a square of side length 1.2 centered on $(-0.1,0)$. We
also examined the effect of viscosity, and a fluid/solid density ratio,
and the scaling of the extra viscous term. The configurations of eight
different tests are shown in Table~\ref{tab:conv_config}.

Due to the complexity of the governing equations, it is near-impossible to
write down an analytical solution to compare against for any test
configuration. We therefore performed reference simulations using a $3960\times
3960$ grid. For each test, we then ran a suite of coarser simulations using
$n\times n$ grids where $n\in \{1980,1320,990,792,660,495,440,396,360\}$ to
compare against the reference results. Since each $n$ divides evenly into
$3960$, the grid squares of these coarse simulations align with the reference
simulations.

\begin{table}
  \begin{center}
    \begin{tabular}{c|c|c|c|c|c|c|c|c}
      Test & State & $\mu$ & $\rho_s$ & CEV & $\vv, L_2$ & $\vv, L_1$ & $\vv, L_\infty$ & $p, L_2$ \\
      \hline
      A & Fluid only & $10^{-3}$ & -- & No & 2.28 & 2.29 & 2.28 & 2.32 \\ %honv0
      B & Fluid only & $4\times 10^{-3}$ & -- & No & 2.32 & 2.32 & 2.32 & 2.33 \\ %honv1
      C & Solid only & $10^{-3}$ & 1 & No & 1.57 & 1.57 & 1.56 & 1.58 \\ %honv10
      C' & Solid only & $10^{-3}$ & 1 & Yes & 2.35 & 2.40 & 1.13 & 2.30 \\ %honv2
      D & Square & $10^{-3}$ & 1 & No & 1.81 & 1.69 & 1.11 & 1.26 \\ %honv11
      E & Circle & $10^{-3}$ & 3 & No & 1.83 & 1.79 & 1.89 & 1.19 \\ %honv12
      F & Circle & $10^{-3}$ & 1 & No & 1.80 & 1.80 & 1.97 & 1.21 \\ %honv13
      F' & Circle & $10^{-3}$ & 1 & Yes & 1.61 & 1.63 & 1.53 & 1.21 %honv5
    \end{tabular}
  \end{center}
  \caption{Details of the eight convergence tests that were performed with
  model problem described in the text. Tests C' and F' were performed using
  constant extra viscosity (CEV) whereby the extra viscosity was held constant
  at the standard value for the lowest resolution grid, $330 \times 330$, as
  opposed to scaling linearly with the grid size. The last four columns give
  the exponents of convergence for velocity $\vv$ and pressure $p$ under
  different $L_q$ norms, based on a linear fit of the three finest-resolution
  data points in Fig.~\ref{fig:conv}.\label{tab:conv_config}}
\end{table}

\begin{figure}
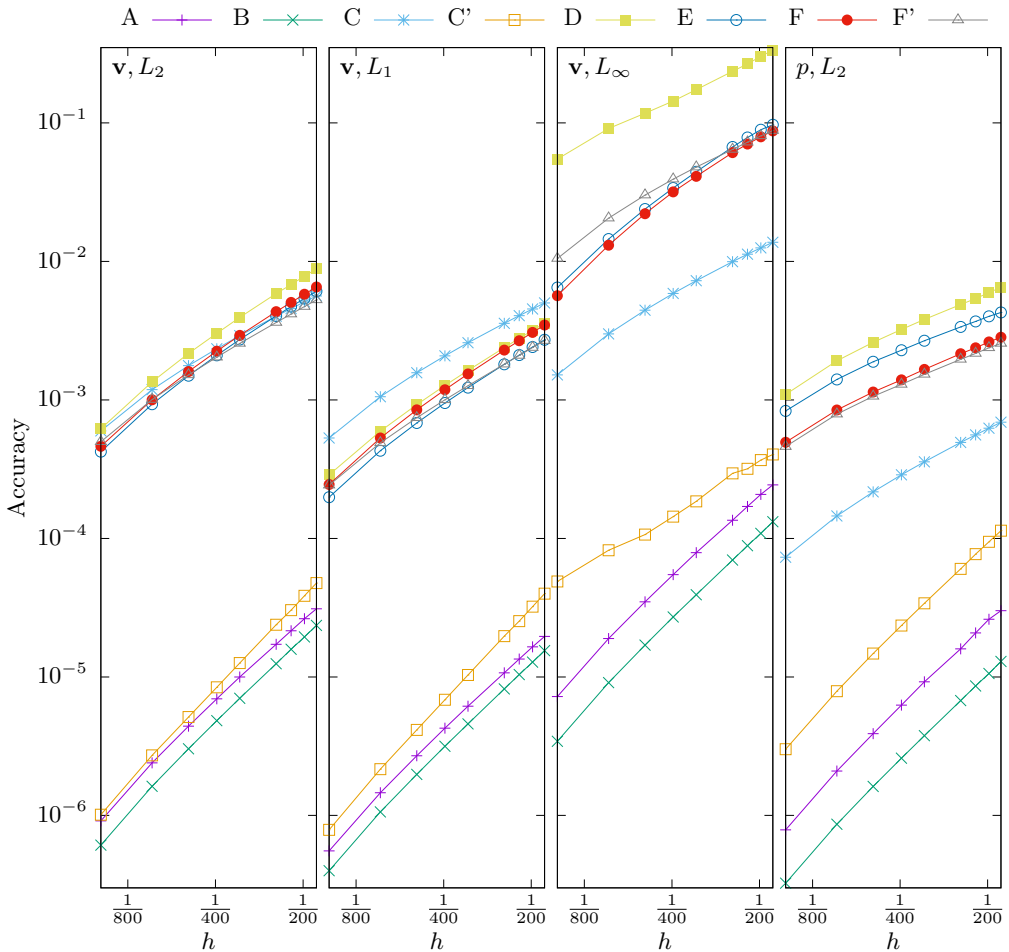

  \begin{center}
    \small
    \include{conv_plot_fig}
  \end{center}
  \vspace{-2.5em}
  \caption{Plots showing the accuracy of the solutions for different grid sizes
  $h$ for the eight convergence tests given in Table \ref{tab:conv_config}.
  Accuracy is computed with respect to reference solutions on a $3960\times
  3960$ grid. Four accuracy measures are shown: the velocity in the $L_2$,
  $L_1$, and $L_\infty$ norms, and the pressure in the $L_2$
  norm.\label{fig:conv}}
\end{figure}

We calculated normalized error measures with respect to $L_q$ norms
\begin{equation}
  E_q^p = \left( \frac{1}{A} \int_\Omega |p_\text{ref}(\vx) - p_\text{coa}(\vx)|^q \,\vec{dx} \right)^{1/q} ,
\quad
E_q^\vv = \left( \frac{1}{A} \int_\Omega ||\vv_\text{ref}(\vx) - \vv_\text{coa}(\vx)||^q_2 \,\vec{dx} \right)^{1/q},
\end{equation}
where $A=4$ is the area of the domain, and the `ref' and `coa' subscripts refer
to the reference and coarse simulation fields, respectively. The integral is
calculated using a direct sum over the field values in the coarser simulation
grid. The pressure field is cell-cornered, and hence each coarse gridpoint
exactly coincides with a reference gridpoint. The velocity field is
cell-centered, so some coarse gridpoints may not align with a reference
gridpoint, in which case the reference value is computed using bilinear
interpolation. The errors associated with this interpolation are $O(h^2)$ and
are small compared to the errors to be measured.

Figure~\ref{fig:conv} shows convergence plots for the velocity in the $L_2$,
$L_1$, and $L_\infty$ norms, plus the pressure in the $L_2$ norm; our
discussion focuses on velocity, since the pressure can be viewed as a Lagrange
multiplier enforcing the incompressibility constraint. For each set of data,
Table \ref{tab:conv_config} lists the corresponding rate of convergence,
calculated using linear regression for the data from the three finest grids,
$n\in \{1980,1320,990\}$. The fluid-only tests, A \& B, are the most accurate,
exhibiting clear second-order convergence across all metrics. Results for the
solid-only test C are less accurate with error measures on the scale of
$10^{-3}$. However, this test is substantially more challenging since the limit
involves tracking elastic waves with zero dissipation. The velocity fields
converge at order $1.6$ in the $L_2$, $L_1$, and $L_\infty$ norms.

Test C uses the usual procedure (Subsec.~\ref{sub:param_stab}) for choosing
extra viscosity, whereby it scales linearly with the grid size. This procedure
is consistent with standard numerical schemes; for example, in the second-order
Lax--Wendroff method~\citep{lax60,leveque_fv} the stabilizing diffusive term
scales linearly with the grid spacing. However, we also considered an
alternative test C' whereby the extra viscosity was chosen based on the
$360\times360$ grid and then held constant for the higher-resolution tests.
This resulted in solutions that were almost as accurate as the fluid-only
tests, with clear second-order convergence in the $L_2$ and $L_1$ norms.
However, the convergence rate in the $L_\infty$ norm is reduced. Inspection of
the results shows that that the maximum deviations are localized to a
one-dimensional set of points where the velocity components are switching sign,
thus resulting in a discrete switch in the timestep update and a lower
convergence rate when measured in the $L_\infty$ norm.

The remaining tests, D, E, F, and F' all involve fluid--structure interaction.
For these tests, the rate of convergence is approximately $1.8$ in the $L_1$
and $L_2$ norms. Inspection of the results shows that the largest deviations
occur at the fluid--structure interface. Since the blur zone is defined in
terms of grid points, its width shrinks at higher resolution. This involves
altering the underlying equations over a region of size $O(h)$, this results in
a lower rate of convergence. However, since the fluid and solid discretizations
are independently second order, is likely that an alternative boundary
treatment---perhaps using a sharp interface approach \citep{gibou05}---could
yield improved results. Test E shows that a fluid--solid density ratio has
little effect on the convergence rate. Test D shows that the square geometry
does not affect the convergence rate in the $L_2$ and $L_1$ norms, but does
result in first-order convergence in the $L_\infty$ norm due to localized
effects at the corners.

\subsection{Flag flapping}
Besides numerical convergence, as a test of the robustness of our approach and
its accuracy across Reynolds numbers, we consider the example of flag flapping,
a problem that has been studied extensively from experimental, numerical, and
analytical perspectives \citep{zhang00,watanabe02,zhu02,connell07}. Following
the problem description and notation of \citet{connell07},\footnote{For
consistency with \citet{connell07} we fully adhere to their notation. However
we draw attention to the reader that two symbols used in this section, $\mu$
(mass ratio) and $h$ (filament thickness), have different meanings than $\mu$
(dynamic viscosity) and $h$ (grid spacing) that are used in all other
sections.} we introduce a thin filament of length $L$, thickness $h\ll L$,
density $\rho_s$, and Young's Modulus $E$, clamped at its left endpoint and
submerged in fluid of kinematic viscosity $\nu$ and density $\rho_f$, flowing
rightward with speed $V$ at infinity. Three dimensionless numbers can be
introduced to study the dynamical behavior of the filament: the mass ratio
$\mu=\rho_sh/\rho_fL$, Reynolds number $\Re=V L/\nu$, and nondimensional
bending rigidity $K_B=EI/(\rho_f V^2 L^3)$. Unlike the previous numerical
approaches that consider the filament to be a one-dimensional beam, our method
uses a true continuum solid formulation so we can consider cases beyond the
thin filament limit, such as a thick flag for which the parameter $h$ does not
necessarily satisfy $h\ll L$.

We first seek to determine if our method correctly captures the transition of
the filament dynamics from stable to flapping in the limit of a thin filament.
We consider a filament with $L=1$, $h=0.05$, $K_B=0.001$, and $\rho_f=1$. To
set $K_B$, we use the fact that in the linear elastic regime $E=3G$, and the
moment of inertia is $I=h^3/12$. We vary $\nu$ and $\rho_s$ in order to test a
range of $\mu$ and $\Re$. The simulation domain is set to be a
$[-1,5]\times[-1,1)$ rectangle with assigned rightward velocity of speed $V=1$
on the left boundary, vanishing pressure on the right boundary, and periodic
boundary conditions on the top and bottom boundaries. We use a $1824\times 608$
grid to represent the domain. The filament is modeled as rectangle $0<x<1,
-\nicefrac{h}{2} < y < \nicefrac{h}{2}$ with semicircular end caps. The
filament is clamped at $(0,0)$ using the tethering methodology described in
Sec.~\ref{spin_rotor}, with $\theta(t)=0$ and $r_\text{teth}=h/4$ in this case.
We track the filament tip by introducing a tracer $\vx(t)$ that starts from
$(1,0)$ and is integrated according to Eq.~\eqref{eq:tracer}. To prevent
integration errors building up over time, the position of the tracer is
periodically reset to satisfy $\vxi_\text{bic}(\vx(t))=(1,0)$ using a
Newton--Raphson root-finding method, where $\vxi_\text{bic}$ is the bicubic
interpolant of the reference map field. The results of this section are based
upon 556 simulations with different parameters that were run on a variety of
Linux and Apple servers at Harvard University and the Lawrence Berkeley
National Laboratory. Depending on parameters and computer speed, each
simulation took been approximately 3--12 days using 4--6 threads. Simulations
with smaller $\Re$ generally take longer, since resolving the fluid viscosity
requires a smaller timestep.

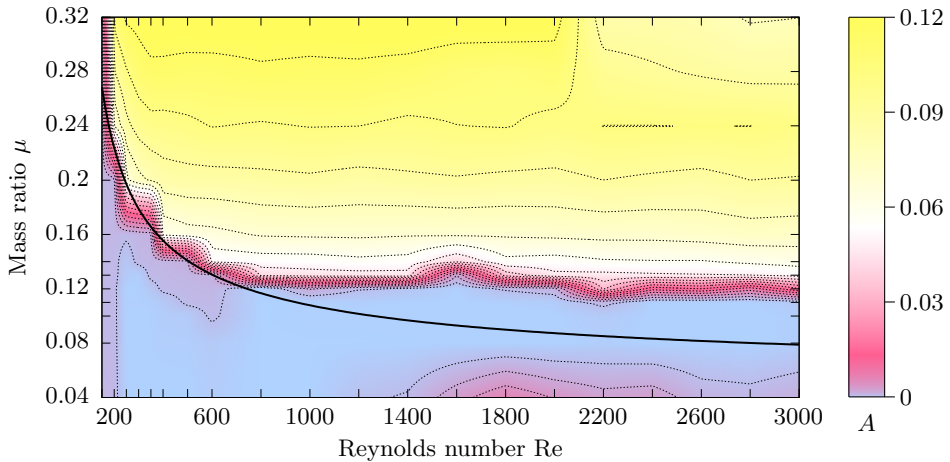
\begin{figure}
  \begin{center}
    \begin{picture}(399,200)
      \put(1,100){\parbox{\textwidth}{\include{tf_pplot_fig}}}
      \put(348,40){\includegraphics[width=14\unitlength,height=148\unitlength]{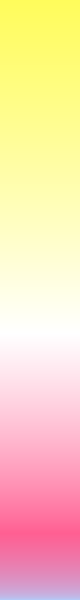}}
      \put(348,40){\line(0,1){148}}
      \put(362,40){\line(0,1){148}}
      \put(348,40){\line(1,0){18}}
      \put(348,188){\line(1,0){18}}
      \put(362,77){\line(1,0){4}}
      \put(362,114){\line(1,0){4}}
      \put(362,151){\line(1,0){4}}
      \put(368,40){\makebox(0,0)[l]{$0$}}
      \put(368,77){\makebox(0,0)[l]{$0.03$}}
      \put(368,114){\makebox(0,0)[l]{$0.06$}}
      \put(368,151){\makebox(0,0)[l]{$0.09$}}
      \put(368,188){\makebox(0,0)[l]{$0.12$}}
      \put(355,30){\makebox(0,0)[c]{$A$}}
    \end{picture}
  \end{center}
  \vspace{-2em}
  \caption{Plot showing the steady-state oscillation amplitude $A$ of a thin flag
  with aspect ratio 20 and bending rigidity $K_B=0.001$, as a function of the
  Reynolds number $\Re$ and mass ratio $\mu$. The colors shown are based on a
  bilinear interpolation of a two-dimensional grid of simulations. The axis ticks
  show the sampled values of $\Re$ and $\mu$, with more simulations
  being performed in parameter ranges of interest. The thin dotted lines are
  contours at spacings of \smash{$\left(\nicefrac{n}{50}\right)^2$} for $n \in
  \mathbb{N}$. The thick solid line is the stable-to-flapping transition
  formula \eqref{eq:connell} of \citet{connell07}.\label{fig:tf_pplot}}
\end{figure}

To systematically evaluate the behavior of the filament, we take the Fourier
transform of the perpendicular deflection of the filament tip over $t\in
[120,160]$, and output the maximum Fourier amplitude, $A$. If $A=0$ the
filament is in the stable (no-flapping) regime and otherwise the filament is
flapping, with $A$ serving as a scalar measure of the amplitude of the dominant
flapping mode. Since our initial conditions are symmetric, the breakage of
symmetry occurs due to numerical noise introduced by the multigrid solver, on
the scale of the parameters $T_\text{MAC}$ and $T_\text{FEM}$ introduced
previously. We also investigated explicitly breaking symmetry by applying an
initial perturbation to the perpendicular velocity in the filament tip, but the
calculations of $A$ were insensitive to this. Since the typical filament
oscillation period is approximately 1.7, the simulations correspond to almost
one hundred complete oscillations, which is sufficient time for the oscillation
amplitude to reach steady state. \cite{connell07} proposed an analytical
formula for the stable-to-flapping transition line:
\begin{equation}
  \label{eq:connell}
  \mu=\frac{1.3 \Re^{-1/2}+K_B4\pi^2}{1-0.65\Re^{-1/2}2\pi-0.5K_B8\pi^3}.
\end{equation}
\citet{connell07} validated this equation numerically using a direct
fluid--filament coupling procedure, a procedure that itself was validated
against experiments \citep{zhang00,watanabe02}. In Fig~\ref{fig:tf_pplot} we
show the behavior of $A$ from our numerical simulations together with the
analytical phase boundary above. For Reynolds numbers below 1000, there is very
good agreement between the locus where $A$ goes non-zero and the analytical
curve. When $\Re\ge1000$ the transition predicted by the simulation happens at
a slightly higher $\mu$ than predicted by Eq.~\eqref{eq:connell}. The most
likely explanation for this is that numerical diffusion from the fluid
advection effectively increases the fluid viscosity. However, other factors
such as the finite domain size, the extensibility of the filament, and the
non-zero $h$ may also play a role.

\begin{figure}
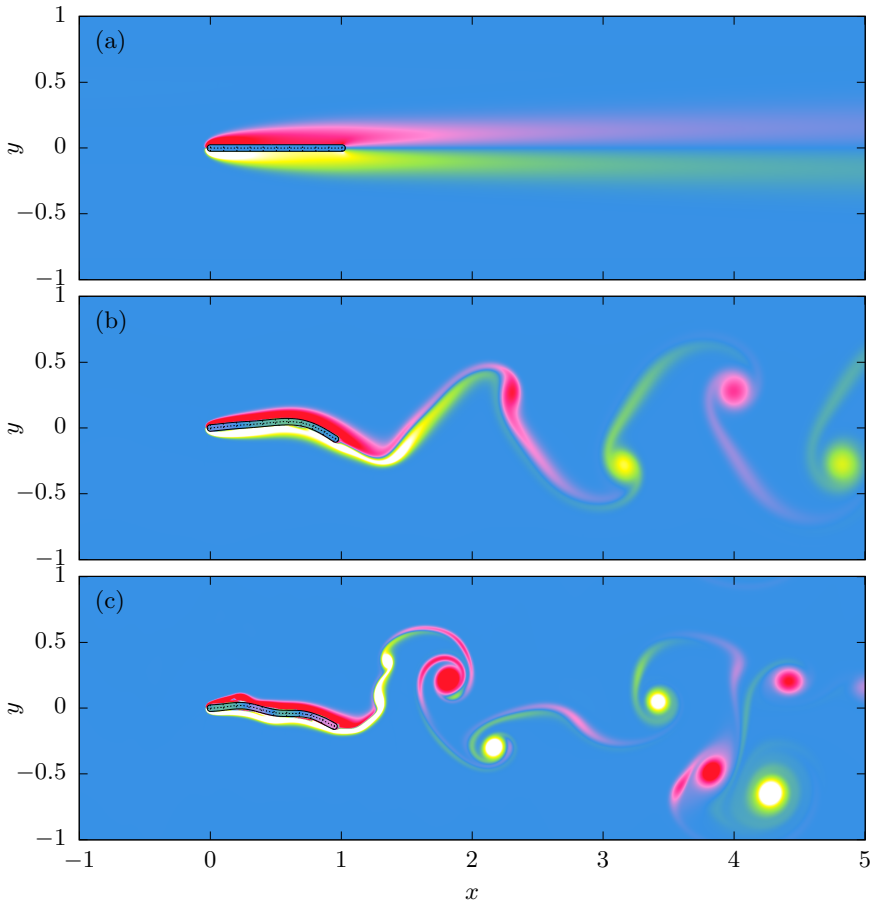

  \begin{center}
    \include{thin_flag_fig}
  \end{center}
  \vspace{-2.4em}
  \caption{Simulations of a thin flexible flag anchored at $(0,0)$ in a fluid
  with mean velocity $\vv=(0,1)$, at $t=160$. The flag has an aspect ratio of
  20. Three simulations with different parameters are shown: (a) stable with
  $(\mu,K_B,\Re)=(0.04,0.001,400)$, (b) limit-cycle flapping with
  $(\mu,K_B,\Re)=(0.16,0.001,1400)$, and (c) chaotic flapping with
  $(\mu,K_B,\Re)=(0.32,0.001,3000)$. The thick black lines mark the
  fluid--structure interfaces. The thin dashed lines are contours of the
  components of the reference map and indicate how the flags deform. The
  colors show vorticity, using the same scale as
  Fig.~\ref{fig:seven-star}.\label{fig:thin_flag}}
\end{figure}

The behavioral switch from stable to flapping is also quite evident in the
long-time flow fields, shown in Fig.~\ref{fig:thin_flag}. Small values of $\mu$
and $K_B$ result in stable behavior, characterized by a straight filament and
fluid flow that is symmetric about the filament axis
(Fig.~\ref{fig:thin_flag}(a)). Upon crossing the transition, periodic
undulatory filament motions develop with a fluid vortex street shed from the
filament (Fig.~\ref{fig:thin_flag}(b)). Increasing $\Re$ and $\mu$ even further
reveals a chaotic filamentary motion, which was also observed in
\citet{connell07} (Fig.~\ref{fig:thin_flag}(c),
\smov{2}{fw_0.32_0.001_3000_20_1e-2}). The chaotic regime coincides with a drop
in $A$ shown in the top right of Fig.~\ref{fig:tf_pplot} because the tip
deflection no longer has a clear single dominant oscillatory mode. Because the
filament is modeled as a thin continuum body of isotropic elastic media as
opposed to an inextensible beam, we observe filament extension in the initial
moments of the simulation as the imposed fluid flow applies a net rightward
traction.

\begin{figure}
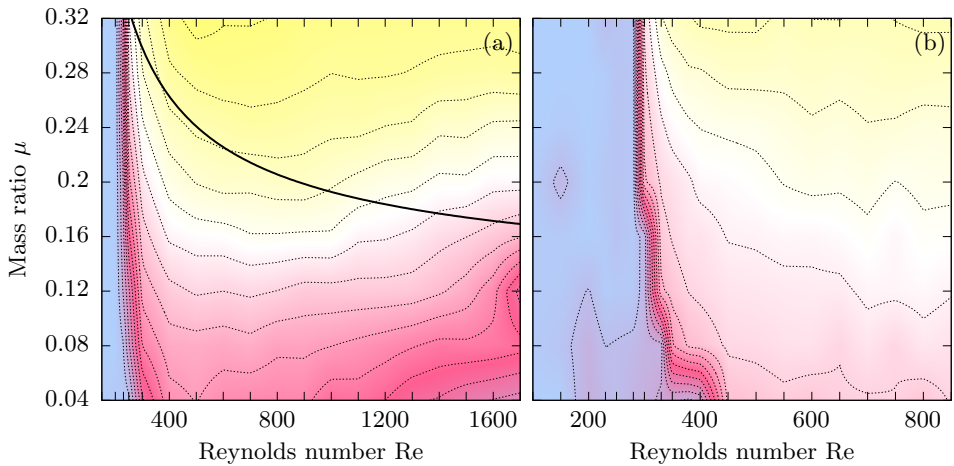

  \begin{center}
    \include{bf_pplot_fig}
  \end{center}
  \vspace{-2em}
  \caption{Plot showing the steady-state oscillation amplitude $A$ as a function of
  the Reynolds number $\Re$ and mass ratio $\mu$, for (a) flags with
  $K_B=0.002$ and aspect ratio 10, and (b) flags with $K_B=0.004$ and aspect
  ratio 5. The colors shown are based on a bilinear interpolation of a
  two-dimensional grid of simulations, using the same scale as
  Fig.~\ref{fig:tf_pplot}. The axis ticks show the sampled values of $\Re$ and
  $\mu$, with more simulations being performed in parameter ranges of interest.
  The thin dotted lines are contour at spacings of
  \smash{$\left(\nicefrac{n}{50}\right)^2$} for $n \in \mathbb{N}$. The thick
  solid line in (a) is the stable-to-flapping transition formula
  \eqref{eq:connell} for thin flags of \citet{connell07}. For (b), the formula
  is out of range and the entire parameter space is in the stable
  region.\label{fig:bf_pplot}}
\end{figure}

\begin{figure}
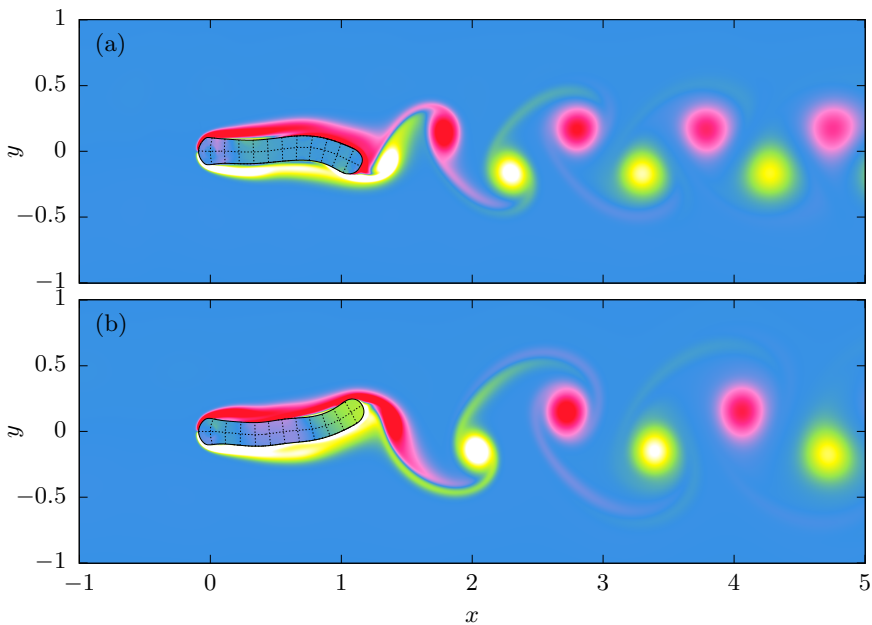

  \begin{center}
    \include{bulky_flag_fig}
  \end{center}
  \vspace{-2.4em}
  \caption{Simulations of a thick flexible flag anchored at $(0,0)$ in a fluid
  with mean velocity $\vv=(0,1)$, at $t=160$. The flag has an aspect ratio of
  5. Two simulations with different parameters are shown: (a) vortex-shedding
  with $(\mu,K_B,\Re)=(0.04,0.004,750)$ and (b) limit-cycle flapping with
  $(\mu,K_B,\Re)=(0.28,0.004,750)$. The thick black lines mark the
  fluid--structure interfaces. The thin dashed lines are contours of the
  components of the reference map and indicate how the flags deform. The colors
  show vorticity, using the same scale as
  Fig.~\ref{fig:seven-star}.\label{fig:bulky_flag}}
\end{figure}

We explore the importance of aspect ratio by introducing $R=h/L$ as an
independent dimensionless group. We observe that as one departs from the $R\ll
1$ regime, adherence to \eqref{eq:connell} is diminished. In
Fig.~\ref{fig:bf_pplot} we show results for $R=10$ and $5$. In general, thick
flags have a smaller stable domain than would be predicted by the thin-filament
limit. We can explain this effect at least in part with bluff-body dynamics.
When $R$ is non-negligible, the thickness of the flag allows the solid geometry
to act as a bluff body over which the fluid is driven to flow. Flow over a
fixed cylinder of diameter $D$ undergoes a transition from a laminar flow to a
periodic vortex street as $DV/\nu$ grows beyond $\sim50$ \citep{lienhard66}. In
our case, the flag thickness acts like $D$, and once a vortex street is induced
off the bluff back end of the flag, the oscillatory force it induces
necessitates flapping. We reiterate that this physical source of oscillatory
forcing emerges only when flags are thick enough to act as a bluff body.
Consistent with this expectation, when $Vh/\nu=\Re\times R>50$ we see only
flapping states for any choice of $\mu$ or $\Re$. Figures
\ref{fig:bulky_flag}(a) and \ref{fig:bulky_flag}(b) show simulation snapshots
of bulky flags with low and high mass ratios, respectively. Simulations of
these two cases are shown in \smov{3}{fw_0.04_0.004_750_5_1e-2} and
\smov{4}{fw_0.28_0.004_750_5_1e-2}, respectively.

\subsection{Solid actuation}
\begin{figure}
  \begin{center}
    \begin{picture}(350,385)
      \put(8,190){\parbox{\textwidth}{\include{rmap1_fig}}}
      \put(314,130){\includegraphics[width=14\unitlength,height=148\unitlength]{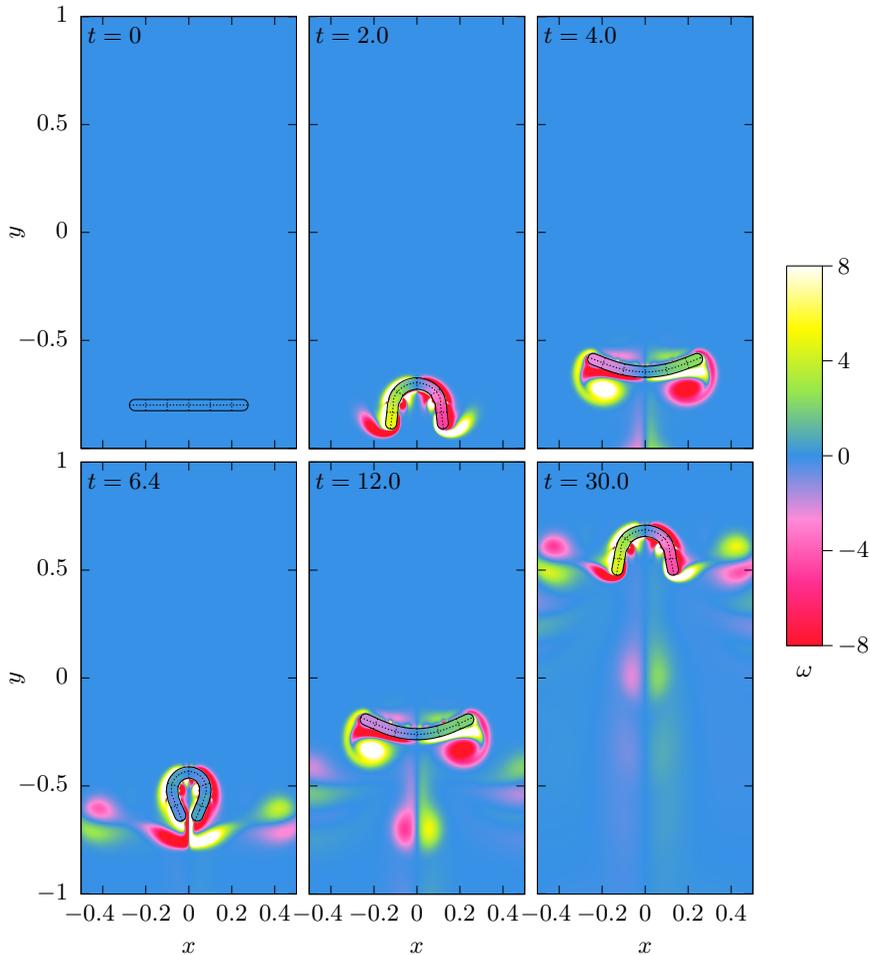}}
      \put(314,130){\line(0,1){148}}
      \put(328,130){\line(0,1){148}}
      \put(314,130){\line(1,0){18}}
      \put(314,278){\line(1,0){18}}
      \put(328,167){\line(1,0){4}}
      \put(328,204){\line(1,0){4}}
      \put(328,241){\line(1,0){4}}
	  \put(334,130){\makebox(0,0)[l]{$-8$}}
	  \put(334,167){\makebox(0,0)[l]{$-4$}}
	  \put(334,204){\makebox(0,0)[l]{$0$}}
	  \put(334,241){\makebox(0,0)[l]{$4$}}
	  \put(334,278){\makebox(0,0)[l]{$8$}}
      \put(321,120){\makebox(0,0)[c]{$\omega$}}
    \end{picture}
  \end{center}
  \vspace{-0.8em}
  \caption{Six successive snapshots of the flapping swimmer ($\Re\approx200$),
  with colors showing vorticity $\omega$. A subregion within the solid body is
  actuated to bend periodically and the remaining solid is passive. The motion
  induces the flapping body to swim. The thick black lines mark the
  fluid--structure interface. The thin dashed lines are contours of the
  components of the reference map and indicate how the swimmer
  deform.\label{fig:flapper}}
\end{figure}

The method also admits a simple approach for simulating actuated solids. This
feature allows one to assign time-dependent internal deformations to subregions
of a solid, which is useful for modeling active media such as swimmers. Unlike
the tethering approach used in Section \ref{spin_rotor}, which assigns the full
motion of a region by adding an external body force in that region, here what is
done is to add extra internal stresses to achieve a desired shape change in a
subdomain, without adding net external force. To actuate a particular
(Lagrangian) solid region, $B_a$, one writes the actuated deformation gradient
$\vF_a(\vX\in B_a,t)$, which can then be equivalently expressed in Eulerian
frame as $\vF_a(\vX=\vxi_a(\vx\in b_a,t),t)$ for $b_a$ the image of $B_a$ in
the Eulerian frame. At any point $\vx\in b_a$, the constitutive relation is
adjusted by replacing all references to $\vF(\vx,t)$ with
$\vF(\vx,t)\vF_a(\vx,t)^{-1}$. In an isotropic hyperelastic system, for
example, this effectively distorts the region's rest configuration to the
distortional state given by $\vF_a$. If at any moment in time a configuration
of the actuated domain differs from the intended actuated configuration, a
stress given by $\mathbf{f}(\vF(\vx,t)\vF_a(\vx,t)^{-1})$ emerges that moves
the system toward the actuated deformation. One could in principle assign a
stiffer response in the actuated domain if a faster conformation is desired,
but we have found it to be sufficient to use the same underlying hyperelastic
constitutive model in the actuated and passive subregions of the solid. This
approach is similar to the multiplicative Kroner--Lee decomposition used in
plasticity \citep{kroner60,lee69}, where a tensorial state variable $\vF_p$ is
introduced and the elastic deformation gradient is given by $\vF\vF_p^{-1}$. But
unlike $\vF_p$, which evolves under a constitutive flow rule, here we assign
$\vF_a(\vx,t)$ directly.

As an example, we consider a flapping swimmer (Fig.~\ref{fig:flapper},
\smov{5}{flapper}). The swimmer is a rectangle of width $W=0.5$ and height
$H=0.052$ with circular end caps, initially centered on $(0,-0.8)$, which we
choose to be the location of the origin. We choose the actuated domain, $B_a$,
to be a centered subregion within the swimmer, comprising a rectangle of width
0.28 and height 0.042 with circular end caps. The following actuation is
applied:
\begin{equation}
  F_a(\vX,t) = \left(
  \begin{array}{cc}
e^{-\alpha(\vX,t)} & 0 \\
0 & e^{\alpha(\vX,t)}
  \end{array}
  \right)
\end{equation}
where
\begin{equation}
  \alpha(\vX,t) = - \lambda X_y H_\epsilon(d) \sin^8 \omega t = - \lambda \xi_y(\vx,t) H_\epsilon(d) \sin^8 \omega t
\end{equation}
and $d$ is the signed distance from the Eulerian boundary of $b_a$. By blurring
the boundary of the actuated domain under $H_\epsilon(d)$, it should be noted
material positioned up to $\epsilon$ away from the true boundary of $b_a$ will
receive some actuation stress. The parameters used in the simulation are
$\omega=2\pi/8$, $\epsilon=2.5 h_x$, and $\lambda=\log (2.2/0.021)$. Thus the
maximum stretch on the top boundary is 2.2. The simulation uses $1200\times
1200$ grid in $[-1.5,1.5)^2$ with periodic boundary conditions.

By actuating the flapper in this fashion, the Lagrangian domain $B_a$, which
comprises roughly half the area of the body, is forced to bend periodically in
time. The unactuated portion of the swimmer remains passive and flaps as an
elastic body in response to be being conjoined to the actuated region. The
swimming flapper achieves a Reynolds
numbers of $\Re=V_{\text{solid}}^{\text{max}}\ W/\nu\sim200$. Its ability to
translate its center of mass by swimming evidences that this example is not
near zero Reynolds number; vortex shedding can be seen for each flap.

\subsection{Multi-body contact}
Since the reference map technique does not employ moving meshes, it is
particularly well-suited to problems involving many objects coming into
contact. This capability would be useful for a variety of problems, such as
studying colloidal mixtures with soft, deformable particles.

To generalize the method to $N$ objects, we introduce independent reference
maps $\vxi^{(1)}, \vxi^{(2)}, \ldots, \vxi^{(N)}$ with the ``$(j)$'' suffix
being used to denote any quantity associated with object $j$. For the purposes
of exposition, we assume each field is defined as a separate globally defined
function that is extrapolated separately, although in reality each reference
map only need be defined in a local neighborhood of each object. Each reference
map is updated using Eq.~\eqref{eq:main_rmap}. For a given $\vxi^{(j)}$, the
solid stress $\vsig_s^{(j)}$ is computed using the methods of
Subsec.~\ref{sub:stress_comp}.

When two or more objects come together, their blur zones may overlap, and thus
it is necessary to generalize the definition of global stress that was given in
Eq.~\eqref{eq:stress_mix}. At a given point, define
$\lambda^{(j)}=1-H_\epsilon(\phi^{(j)})$ to be the solid fraction of object
$j$. Then the stress is given by
\begin{equation}
  \label{eq:stress_mix2}
  \vsig =
  \begin{cases}
    \vsig_{f} + \sum_i \lambda^{(i)} (\vsig_s^{(i)}-\vsig_f)_{\strut} & \qquad \text{if $\sum_i \lambda^{(i)}\le 1$,} \\
    \dfrac{\sum_i \lambda^{(i)} \vsig_s^{(i)}}{\sum_i \lambda^{(i)}} & \qquad \text{if $\sum_i \lambda^{(i)}>1$.}
  \end{cases}
\end{equation}
If only one object is present, this definition exactly matches
Eq.~\eqref{eq:stress_mix}. If several objects are present, then they each
contribute to the global stress, with the fluid stress filling any unassigned
fraction. In rare situations (\textit{e.g.}~three objects meeting at a point)
the solid fractions may total more than one. In this case, $\vsig$ is taken as
a weighted average of the solid stresses, and the fluid stress does not
contribute at all. The global density field is defined using the same mixing
procedure as in Eq.~\eqref{eq:stress_mix2}.

In our tests, we have found that independently updating $N$ reference maps and
computing a global stress according to Eq.~\eqref{eq:stress_mix2} is sufficient
to perform multi-body simulations. However, since the simulation employs a
single globally-defined velocity field, it becomes problematic when shapes
become very close together, since it is hard for them to separate as they move
according to the same underlying velocity. Similar behavior has been noted in
the literature on the immersed boundary method \citep{lim12,krishnan17}, which
also employs a single global velocity field for the movement of structures. To
rectify this, we introduce a small contact stress (in addition to the stress of
Eq.~\eqref{eq:stress_mix2}) when the blur zones of two objects overlap, which
penalizes the interfaces from becoming too close together. We first define a
contact force function of
\begin{equation}
  \label{eq:contact_force}
  f(\alpha) =
  \begin{cases}
    \tfrac{1}{2}(1-\tfrac{\alpha}{\epsilon}) & \qquad \text{if $\alpha<\epsilon$,} \\
    0 & \qquad \text{if $\alpha\ge \epsilon$.}
  \end{cases}
\end{equation}
Now, consider the stress calculation at an edge that is within the blur zones
of two or more solids. Consider a pair of the solids $(i)$ and $(j)$. Using
finite differences, compute a unit normal vector
\begin{equation}
  \nor = \frac{\nabla (\phi^{(i)}-\phi^{(j)})}{|| \nabla(\phi^{(i)}-\phi^{(j)}) ||_2}
\end{equation}
where $||\cdot||_2$ denotes the Euclidean norm. The contact stress is defined
as
\begin{equation}
  \vsig_\text{col} = - \eta \min \{ f(\phi^{(i)},f(\phi^{(j)}) \} (G^{(i)}+G^{(j)})
  (\nor \otimes \nor - \Ident ),
\end{equation}
where $\eta$ is a dimensionless constant, the $G^{(i)}$ are object-dependent
shear moduli, and the $\Ident$ term is included to make the stress trace-free.
In the rare case where the edge is within three or more solid blur zones, the
calculation is repeated $\vsig_\text{col}$ for each pair, and each contribution
is added to the global stress.

These collision stress terms induce forces that push apart objects when they
become close. Formulating the collision interaction as an additional stress is
advantageous since it immediately ensures that total momentum of the entire
simulation is numerically conserved. The method is not sensitive to the exact
functional form of $f$ in Eq.~\eqref{eq:contact_force}. An alternative
formulation is to directly use the transition function,
$f(\alpha)=1-H_\epsilon(\alpha)$, but we find that the faster growth of the
function in Eq.~\eqref{eq:contact_force} when $\alpha$ becomes smaller than
$\epsilon$ yields smoother results in our test cases.

\begin{figure}
  \begin{center}
    \begin{picture}(400,510)
      \put(8,250){\parbox{\textwidth}{\include{multi-drop_fig}}}
      \put(354,190){\includegraphics[width=14\unitlength,height=148\unitlength]{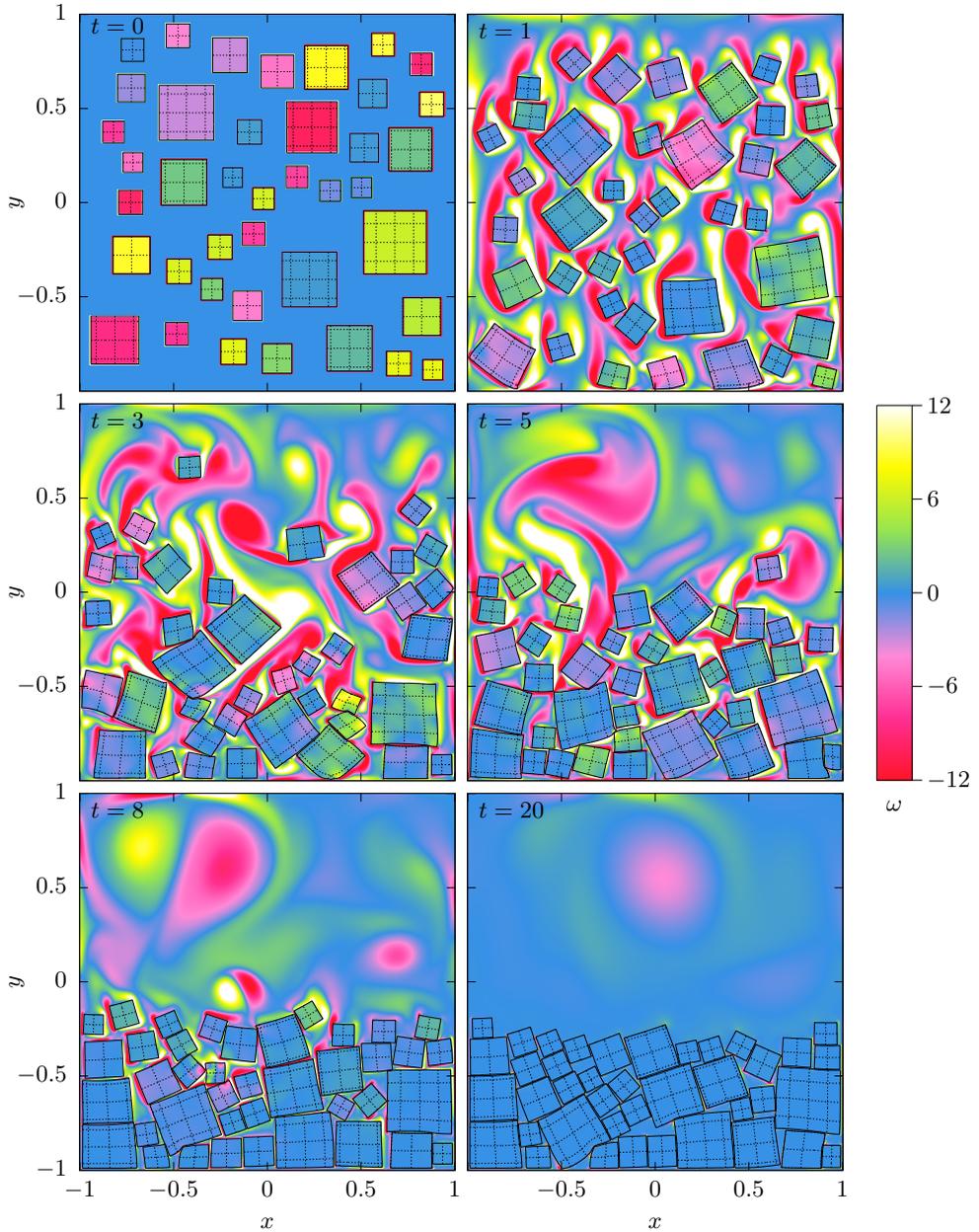}}
      \put(354,190){\line(0,1){148}}
      \put(368,190){\line(0,1){148}}
      \put(354,190){\line(1,0){18}}
      \put(354,338){\line(1,0){18}}
      \put(368,227){\line(1,0){4}}
      \put(368,264){\line(1,0){4}}
      \put(368,301){\line(1,0){4}}
	  \put(374,190){\makebox(0,0)[l]{$-12$}}
	  \put(374,227){\makebox(0,0)[l]{$-6$}}
	  \put(374,264){\makebox(0,0)[l]{$0$}}
	  \put(374,301){\makebox(0,0)[l]{$6$}}
	  \put(374,338){\makebox(0,0)[l]{$12$}}
      \put(361,180){\makebox(0,0)[c]{$\omega$}}
    \end{picture}
  \end{center}
  \vspace{-1em}
  \caption{Snapshots of vorticity $\omega$ in a simulation of forty squares
  sedimenting in a fluid-filled box. The thick black lines mark the
  fluid--structure interfaces. The thin dashed lines are contours of the
  components of the reference map defined in each object and indicate how the
  squares deform.\label{fig:multi-drop}}
\end{figure}

Figure~\ref{fig:multi-drop} shows snapshots from a multi-body simulation in a
non-periodic box $[-1,1]^2$ using a resolution of $1000\times 1000$.Forty
squares with shear modulus $G=2$ and density $\rho_s=3$ are inserted at random
positions in the box, with side lengths chosen uniformly over the range
$[0.1,0.4]$. Any squares that lie within a distance of $0.1$ of another square
are rejected, and are chosen again. At $t=0$, each square is set to initially
spin with angular velocity chosen uniformly from the range $[-5,5]$. A
gravitational acceleration of 0.5 in the negative $y$ direction is applied, so
that the squares sediment at the bottom of the box. The full simulation is
shown in \smov{6}{multi-drop}.

\section{Conclusion}
Herein, we have presented a robustly accurate, yet straightforward to
implement, reference map technique, which has allowed us to study a variety of
FSI problems using a single background grid. It augments the multi-phase fluid
framework of \citet{yu03} by allowing general finite-deformation solid models to
be coupled directly to a fluid. In doing so, it maintains a number of the
advantages of working on a fixed Eulerian grid that are enjoyed in fluid
simulation methods. The practicality and usefulness of this approach is
demonstrated in various tests. It is shown to capture the flapping phase
diagram for thin flags and the transition from thin- to thick-flag behaviors,
which highlights the role of new mechanisms to initiate flapping. Additional
physics, such as actuation of solids, is straightforward to implement with a
user-described actuated deformation gradient. This capability is used to model
a swimming object with realistic internal driving. The ability to model objects
with sharp corners is typically a challenge in Eulerian approaches, but here it
can be done by exploiting the reference map field near the edge of the object.
We also present an improved contact algorithm, which we use to simulate
situations with many soft interacting objects submerged in a fluid.

There are a number of future directions. One of clearest applications is in
biomechanics, with the simulation of systems of many interacting, actuated
cells. We also foresee modeling solids beyond hyperelasticity, such as
plasticity, thermal material models, and growth. These modifications can be
done through the inclusion of new state variables in the solid and/or the
addition of a heat diffusion equation; there are clear advantages to
implementing thermal diffusion in the Eulerian frame. Beyond extensions to
three dimensions, there are opportunities to use the approach for
dimensionally-reduced models such as membranes and shells by restricting the
reference map to a lower dimensional set. Regarding contact modeling, the
reference map field could be used to instruct formulations for more advanced
contact problems, including friction and self-contact. Lastly, it is a major
goal to extend the approach to allow for non-persistent material boundary sets,
as occurs in fracture. It may be possible to represent crack surfaces through
intersecting level set fields and to couple this capability with physical
traction--separation relations to generate new surface material as cracks
advance.

\appendix
\section{Additional numerical details}
\subsection{Monotonicity-limited derivative}
\label{app:mono_deriv}
The gradients of the reference map and velocity appearing in
Eq.~\ref{eq:godunov1} are computed using the fourth-order monotonicity-limited
scheme of \citet{colella85}. For the derivative of a generic component
$\gener_{i,j}$ the $x$ direction, finite differences
\begin{equation}
  D^c(\gener)_{i,j} = (\gener_{i+1,j} - \gener_{i-1,j})/2,
  \quad D^+(\gener)_{i,j} = \gener_{i+1,j} - \gener_{i,j},
  \quad D^-(\gener)_{i,j} = \gener_{i,j} - \gener_{i-1,j}
\end{equation}
are introduced, from which the limiting slope is defined as
\begin{equation}
  \delta_\text{lim}(\gener)_{i,j} =
  \begin{cases}
    2\times \min(|D^-(\gener)_{i,j}|,|D^+(\gener)_{i,j}|) & \qquad \text{if $D^-(\gener)_{i,j}D^+(\gener)_{i,j}>0$,} \\
    0 & \qquad \text{otherwise.}
  \end{cases}
\end{equation}
The second-order limited slope is then
\begin{equation}
  \delta_f(\gener)_{i,j} = \min (|D^c(\gener)_{i,j}|,\delta_\text{lim}(\gener)_{i,j})\times \sign(D^c(\gener)_{i,j})
\end{equation}
from which the fourth-order monotonicity limited derivative is defined as
\begin{equation}
  \delta^4(\gener)_{i,j} = \min \left(\frac{|8 D^c(\gener)_{i,j} - \delta_f(\gener)_{i+1,j} - \delta_f(\gener)_{i-1,j}|}{3},\delta_\text{lim}(f)_{i,j} \right)
  \times \frac{\sign(D^c(f)_{i,j})}{h_x}.
\end{equation}
The $y$-derivative is evaluated similarly.

\subsection{Tangential derivatives}
\label{app:tang}
To ensure stability, the tangential derivatives appearing in
Eqs.~\eqref{eq:godunov1} \& \eqref{eq:godunov2} are computed using
\begin{align}
  (\widetilde{v \vxi_y})^n_{i,j} &= \frac{\tilde{v}^\adv_{i,j-1/2} + \tilde{v}^\adv_{i,j+1/2}}{2}
  \frac{\tilde{\vxi}_{i,j+1/2} - \tilde{\vxi}_{i,j-1/2}}{h_y}, \label{eq:tderiv1} \\
  (\widetilde{v \vv_y})^n_{i,j} &=\frac{\tilde{v}^\adv_{i,j-1/2} + \tilde{v}^\adv_{i,j+1/2}}{2}
  \frac{\tilde{\vv}_{i,j+1/2} - \tilde{\vv}_{i,j-1/2}}{h_y}, \label{eq:tderiv2}
\end{align}
where the terms with tildes are computed using a preliminary Godunov upwinding
step where stress, pressure, and tangential derivatives are neglected
\citep{yu03}. Extrapolations to the right edge are given by
\begin{align}
  \tilde{\vxi}^{R,n+1/2}_{i+1/2,j} &= \vxi_{i,j} + \frac{1}{2} \left( h_x - u^n_{i,j} \Delta t\right) \vxi^n_{x,i,j}, \\
  \tilde{\vv}^{R,n+1/2}_{i+1/2,j} &= \vv^n_{i,j} + \frac{1}{2} \left(h_x - u^n_{i,j} \Delta t\right) \vv^n_{x,i,j},
\end{align}
and with extrapolations to the other edges given similarly. On each edge, with
the selection procedure of Eq.~\eqref{eq:gselect} is used, with Case A used for
$\tilde{\vv}^\adv=(\tilde{u}^\adv,\tilde{v}^\adv)$ and Case B used for
$\tilde{\vxi}$ and $\tilde{\vv}$.

\section*{Acknowledgments}
C.~H.~Rycroft was supported by the Director, Office of Science, Computational
and Technology Research, U.S.~Department of Energy under contract number
DE-AC02-05CH11231. Y.~Yu acknowledges support from the National Science
Foundation under Award No.~DMS-1620434 and the Class of 1968 Junior Faculty
Fellowship from Lehigh University.

\bibliography{refs}

\end{document}